\documentclass[%
reprint,
superscriptaddress,
 amsmath,amssymb,
 aps,
 prf,
]{revtex4}

\usepackage{graphicx}
\usepackage{dcolumn}
\usepackage{bm}
\usepackage{color}
\usepackage{soul}
\usepackage{array}
\newcommand{\ub}{\mathbf{u}}
\newcommand{\ml}[1]{{{\color{black}#1}}}
\newcommand{\vc}[1]{{{\color{black}#1}}}
\newcommand{\vk}[1]{{{\color{black}#1}}}

\newcommand{\mkl}[1]{{{\color{black}#1}}}

\begin{document}

\preprint{APS/123-QED}

\title{\vc{R}econstruct\vc{ing} the time evolution of wall-bounded turbulent flows from non-time resolved PIV measurements}

\author{C. Vamsi Krishna}
\email{vchinta@usc.edu}
\affiliation{Aerospace and Mechanical Engineering,\\
University of Southern California, Los Angeles, CA 90089}

\author{Mengying Wang}
\affiliation{
 Aerospace Engineering and Mechanics,\\
 University of Minnesota, Minneapolis, MN 55455
}%


\author{Maziar S. Hemati}
\affiliation{
 Aerospace Engineering and Mechanics,\\
 University of Minnesota, Minneapolis, MN 55455
}%

\author{Mitul Luhar}
\affiliation{Aerospace and Mechanical Engineering,\\
University of Southern California, Los Angeles, CA 90089}



\begin{abstract}
Particle Image Velocimetry (PIV) systems are often limited in their ability to fully resolve the spatiotemporal fluctuations inherent in turbulent flows due to hardware constraints.  In this study, we develop 
models based on Rapid Distortion Theory (RDT) and Taylor's Hypothesis (TH) to reconstruct the time evolution of a turbulent flow field in the intermediate period between consecutive PIV snapshots obtained using a non-time resolved system.  The linear governing equations are \ml{evolved} forwards and backwards in time using the PIV snapshots as initial conditions.  The flow field in the intervening period is then reconstructed by taking a weighted sum of the forward and backward estimates.  This spatiotemporal weighting function is designed to account for the advective nature of the RDT and TH equations.  Reconstruction accuracy is evaluated as a function of spatial resolution and reconstruction time horizon using Direct Numerical Simulation data for turbulent channel flow from the Johns Hopkins Turbulence Database. This method reconstructs single-point turbulence statistics well and resolves velocity spectra at frequencies higher than the temporal Nyquist limit of the acquisition system.  
\ml{Reconstructions obtained using a characteristics-based evolution of the flow field under TH prove to be more accurate compared to reconstructions obtained from numerical integration of the discretized forms of RDT and TH.}
The effect of measurement noise on reconstruction error is also evaluated.
\end{abstract}

\maketitle

\section{Introduction}\label{sec:intro}
\begin{figure}
    \centering
        \includegraphics[trim = {0.5cm 1.5cm 0cm 1.5cm},scale=0.55, clip=true]{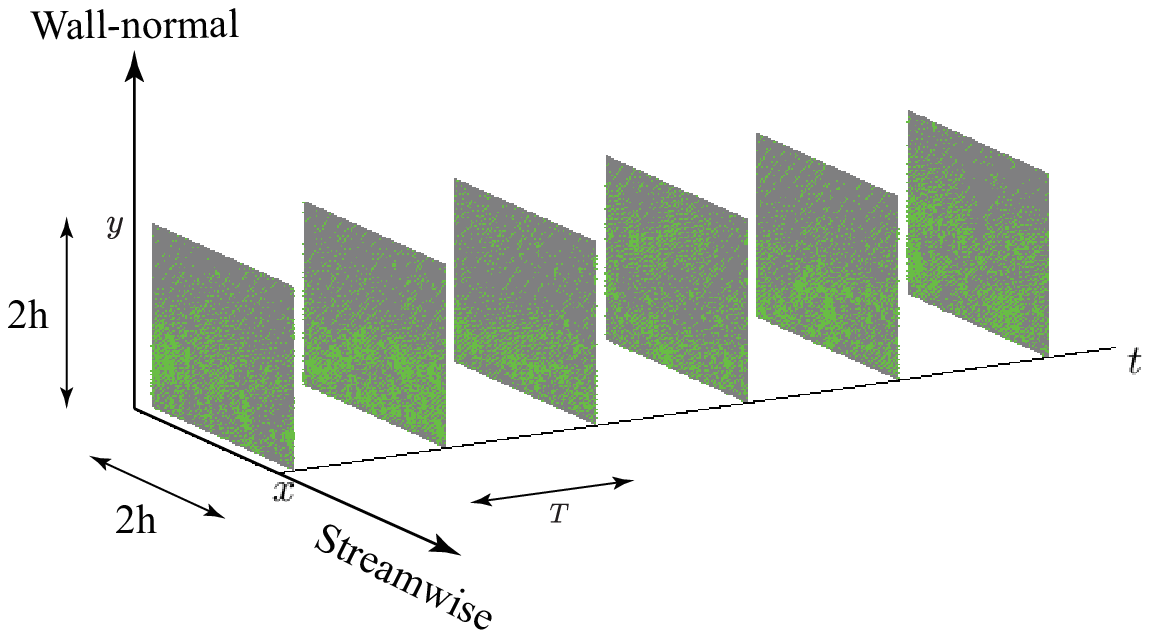}
        \includegraphics[trim = {1cm 0cm 0cm 0.5cm},scale=0.55, clip=true]{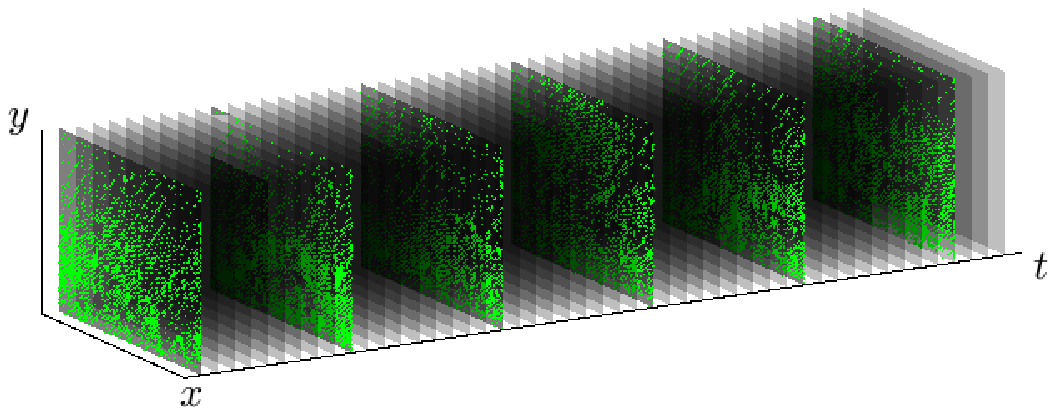}
    \caption{Left: Non-time resolved PIV snapshots stacked along the time axis.  Right: Time-resolved reconstruction of the velocity field from the PIV snapshots.}
    \label{fig:schematic}
\end{figure}

\subsection{Motivation and Problem Statement}
The ability to generate spatially and temporally resolved velocity measurements in turbulent flows is essential \ml{for} improving our understanding of the underlying dynamics, identifying coherent structures, and the development of flow control. \ml{The  last  three  decades  have  seen  rapid advances in the development of high-power and high-repetition rate lasers, high-speed digital cameras capable of megapixel resolution, and computing power.  These hardware advances, together with improvements in the speed and accuracy of Particle Image Velocimetry (PIV) analysis algorithms, have led to a step change in our ability to make non-intrusive field measurements at high spatial and temporal resolution \citep{westerweel2013particle,adrian1991particle}.}
\ml{Despite these advances}, laboratory PIV systems are often limited in their ability to fully resolve the broadband spatiotemporal fluctuations inherent in turbulent flows \ml{due to hardware limitations or cost constraints.  For example}, even state-of-the-art PIV systems with kHz-capable cameras and lasers may not yield complete temporal resolution in certain conditions.  These limitations in hardware motivate the need to reconstruct the flow field from limited and noisy measurements. 

In this study, we develop 
models that can reconstruct the time evolution of wall-bounded turbulent flows in the intervening period between two PIV snapshots from a non-time resolved system.  Figure~\ref{fig:schematic} provides a schematic view of the problem being addressed.  The left panel in Figure~\ref{fig:schematic} shows PIV snapshots stacked along the time axis. Consistent with typical planar PIV systems, we assume that only two-dimensional, two-component (2D-2C) snapshots are available.  The sampling time between PIV measurements is $T$.  As shown in the right panel, the goal is to reconstruct the evolution of the flow field between two consecutive snapshots with high temporal resolution, i.e., to generate predictions for the snapshots shown as translucent planes.  We focus on turbulent channel flow and use direct numerical simulation (DNS) data from the Johns Hopkins Turbulence Database \citep{graham2016web} to develop and test the models used for flow reconstruction. However, we expect these techniques to be equally applicable to other wall-bounded flows (e.g., pipe and boundary layer flows) and be directly transferable to physical PIV systems. 

Most prior efforts on turbulent flow reconstruction have relied on data-driven approaches, in which sensor data are used in conjunction with static correlation maps or projected onto basis functions obtained from field measurements \citep[e.g.,][]{adrian1994stochastic, bonnet1994stochastic, taylor2004towards, tu2013integration,berry2017low, discetti2018estimation}.  In contrast, the present effort employs simplified models based on Rapid Distortion Theory (RDT) and Taylor's Hypothesis (TH) that are derived from the governing Navier-Stokes equations. 
A brief review of previous reconstruction efforts and models grounded in RDT/TH is provided below. 

\subsection{Previous Reconstruction Efforts}
From a signal processing point of view, time resolution issues in measurements can be alleviated through the use of compressed sensing, particularly when the signal of interest is sparse in frequency space \citep[e.g.,][]{candes2008restricted,bai2014low,tu2014spectral}.  If such narrow-banded signals are sampled randomly in time, then $\ell_1$ minimization techniques can be used to reconstruct their time evolution even with sub-Nyquist average sampling rates.  However, such techniques are less reliable for signals that are not sparse in frequency space, as is the case with broad-banded turbulent flows.  Moreover, it is not typically possible to generate PIV measurements that are sampled randomly in time.

Previous studies have compensated \vc{for} the limited time resolution of PIV systems through the use of two complementary instruments.  In this multi-sensor fusion approach, the high-spatial and low-temporal resolution velocity fields from PIV are typically fused with time-resolved point measurements from instruments such as hot-wire anemometers (HWA) or pressure sensors \citep{tu2013integration,berry2017low,discetti2018estimation}. Broadly, such sensor fusion and flow field reconstruction techniques can either involve a static approach, such that the estimator relies purely on statistics compiled from prior data, or a dynamic approach, in which an underlying evolution model is included in the estimation procedure. 

Stochastic estimation with static maps has been employed extensively in the turbulence community \citep[e.g.,][]{adrian1994stochastic, cole1992application, guezennec1989stochastic, murray2007modified}, particularly in conjunction with proper orthogonal decomposition (POD) \citep{bonnet1994stochastic, taylor2004towards}. Variants of Linear Stochastic Estimation (LSE) have been used in many different contexts. For instance, POD-based LSE has been used to educe coherent structure in turbulent flows \citep{adrian1988stochastic, bonnet1994stochastic, guezennec1989stochastic}; multi-sensor \ml{stochastic estimation} has been used to estimate velocity fields from \ml{wall-based pressure and shear stress measurements \citep{murray2003estimation, naguib2001stochastic,encinar2019logarithmic}}; and LSE incorporating time delays has been used to predict the future evolution of flow fields for purposes of control \citep{taylor2004towards,ukeiley2008dynamic}. \ml{Along similar lines, \citet{sasaki2019transfer} have recently used Large Eddy Simulation data to identify linear and nonlinear transfer functions that enable the estimation of streamwise velocity fluctuations in turbulent boundary layers based on measurements at different wall-normal locations.} Of relevance to the present study, LSE has \ml{also} been employed to fuse information from fast-time point measurements and slow-time field data for turbulent channel flow. Specifically, \citet{discetti2018estimation} evaluated the correlation between fast-time point measurements and the time variation in POD modes obtained from the field measurements, at the instances in which these datasets were synchronized. This static correlation map was then used to infer the fast-time evolution of the POD modes from the point measurements. 


\vc{D}ynamic estimators seek to incorporate information from underlying dynamic models as well as sensor measurements to estimate the state of high-dimensional systems such as turbulent flows. 
Standard techniques for linear dynamic estimation include Kalman filtering and Kalman smoothing.  Kalman filtering estimates the current state of the system based on previous observations, while Kalman smoothing also allows for refinement of previous estimates in light of later observations. \citet{tu2013integration} employed a Kalman smoother to successfully fuse information from fast-time point measurements and slow-time field measurements to reconstruct the velocity field in the wake of a thick flat plate with an elliptical leading edge at low Reynolds number. The dynamic model employed in this study involved projection onto POD modes computed from the field measurements. \citet{tu2013integration} found that a dynamic model comprising the following two elements was sufficient for accurate flow field reconstruction: (i) the first two POD modes oscillating stably at the shedding frequency measured from the probe signal, with amplitudes estimated from the point measurements, and (ii) the remaining POD mode coefficients advanced in time via LSE. For the purposes of more general turbulent flow reconstruction, dynamic models like the one developed by \citet{tu2013integration} have two important limitations. First, the assumption of stable, low dimensional oscillatory dynamics only holds for a small class of narrow-banded flows (e.g., low Reynolds number bluff body wakes). Second, projection onto POD modes limits any model predictions to the subspace spanned by the data, and does not guarantee that the resulting flow fields will be physically sound. This limitation is especially problematic in the context of the broadband turbulent flows to be considered in this study. Indeed, the necessary data for extracting the appropriate POD-basis (or a reliable data-driven model by other means) are often unavailable, even when using state-of-the-art instrumentation for flow diagnostics.

To circumvent these issues relating to data availability, recent studies have attempted flow reconstruction by projecting the velocity field onto so-called resolvent modes that are obtained via a gain-based decomposition of the governing equations \citep{mckeon2010critical,mckeon2017engine}. \ml{For instance, this approach has been used to reconstruct the unsteady three-dimensional flow field in a lid-driven cavity flow based on a limited set of point measurements of velocity \citep{gomez2016reduced}; to estimate the pressure distribution around an inclined square cylinder based on time-resolved measurements at a single point in the wake \citep{gomez2016estimation}; to model the flow around an airfoil based on limited PIV measurements \citep{symon2019tale}; and to estimate the cross-spectral density of the turbulent fluctuations in a jet issuing from a convergent-straight nozzle \citep{lesshafft2019resolvent}.  Typically, such resolvent-based reconstruction efforts proceed as follows.  First, the point or field measurements of velocity are used to identify the dominant frequencies present in the flow. Next, a limited set of resolvent modes are computed for these frequencies, and their amplitudes are calibrated using the measurements.  The time-varying flow fields can then be reconstructed based on a linear superposition of the calibrated resolvent modes.  Note that a similar procedure has also been used to reconstruct the flow field for a round jet using modes obtained from analysis of the parabolized stability equations \citep{beneddine2017unsteady}.} Since resolvent \ml{or stability} modes can be computed directly from the governing equations if a base (or mean) velocity profile is available \ml{\citep[e.g., from PIV; see][]{symon2019tale,lesshafft2019resolvent}}, such \ml{equation}-based flow reconstruction \ml{techniques} minimize the need for a priori data.  Moreover, the use of resolvent modes ensures that the reconstructed flow fields will be physically sound (e.g., satisfy the continuity constraint in incompressible flow).  At the same time, this approach is most useful in band-limited flows for which a limited set of modes can serve as an adequate basis for projection.  Identification, computation, and calibration of dominant resolvent modes is much more challenging in broad-banded turbulent flows.  For completeness, we note that \vc{\citet{illingworth2018estimating} have made use of the linearized NSE} to estimate the velocity field at a given wall-normal location within a turbulent channel flow using time-resolved velocity measurements from a different wall-normal location and \vc{\citet{towne2020resolvent} have used the resolvent formulation} to estimate the space-time statistics in \vc{a} turbulent channel flow. 

\subsection{Rapid Distortion Theory and Taylor's Hypothesis}
As noted earlier, in this paper we use models grounded in Rapid Distortion Theory (RDT) and Taylor's hypothesis (TH) to reconstruct the flow field between two consecutive PIV snapshots.  
In essence, RDT assumes that if a turbulent flow field is subjected to substantial distortion by the mean shear flow, the higher-order nonlinear interactions can be neglected when predicting the early response. Scaling arguments show that RDT is formally correct if the time-horizon of prediction is much shorter than a typical eddy turnover time \cite{hunt1990rapid}. This makes RDT a natural choice for the present effort, in which the velocity field must be evolved from a known initial state over a short time horizon until the next PIV snapshot becomes available.  
RDT has been used extensively in both theoretical and experimental turbulence research \ml{\citep{savill1987recent,hunt1990rapid}}.  Moreover, there are strong connections between RDT and the resolvent analysis framework mentioned in the previous section \citep{mckeon2017engine} since both approaches emphasize linear dynamics.

Under Taylor's Hypothesis (TH), the RDT equations are simplified further to retain just the time derivative of the fluctuations and the mean flow advection term.  In other words, TH assumes that the turbulent flow field is `frozen' and advects downstream with the mean flow \citep{taylor1938spectrum}. The mean velocity can therefore be used to convert temporal information to spatial information, and vice versa. TH has been used extensively in the turbulence community \vc{\citep{wyngaard1977taylor,dennis2008limitations,delAlamo2009estimation,moin2009revisiting,yang2018implication}}, primarily to infer spatial structure from time-resolved point measurements (e.g., from hot-wire anemometers).  Experimental estimates of streamwise wavenumber ($k_x$) spectra are often converted from frequency ($\omega$) spectra under the assumption that the resulting convection velocity ($c=\omega/k_x$) is equal to the local mean velocity \citep{smits2011high}.  This is a good assumption in the outer region of the flow, where the convection velocity of the turbulent fluctuations is known to be close to the local mean velocity. However, experiments and numerical simulations both show that this assumption leads to an underestimate of the convection velocity below the buffer region of the flow. Specifically, the convection velocity remains at $c^+ \approx 10$ below $y^+ \approx 15$, even as the mean velocity goes to zero at the wall  \citep[see e.g.,][]{kim1993propagation,krogstad1998convection,quadrio2003integral}. Here, $y$ is the wall-normal coordinate ($y=0$ at the wall) and a superscript $+$ denotes normalization with respect to the friction velocity and kinematic viscosity. Thus, the use of the local mean velocity in TH leads to underestimation of streamwise length scales in the near-wall region. Moreover, wall-bounded turbulent flows are typically characterized by a broad spectrum of frequencies for a given wavenumber (and vice versa). Hence, using the same convection velocity for all wavenumber-frequency combinations at a given wall-normal location can also lead to spurious peaks in the power spectral density \citep{delAlamo2009estimation,moin2009revisiting}.  

Since TH has been used to infer spatial structure from time-resolved measurements with some success in turbulent flows, it should also be possible to invoke this hypothesis to solve the inverse problem: to infer the time evolution of a flow field from measurements of spatial structure. The reconstruction framework described in this paper develops this idea further.

\subsection{Contribution and Outline}
To reconstruct the time evolution of the flow field between consecutive 2D-2C PIV snapshots, we use the simplified equations obtained under RDT and TH to \ml{generate predictions} forwards in time from the initial snapshot and backwards in time from the subsequent snapshot.  A weighted sum of these forward and backward estimates is used to reconstruct the flow field in the intervening period.  The accuracy of the reconstructed flow fields obtained using these models is assessed using DNS data for turbulent channel flow at friction Reynolds number $Re_\tau = 1000$ obtained from the Johns Hopkins Turbulence Database (JHTDB)\citep{graham2016web}. Results show that this 
reconstruction framework significantly outperforms direct interpolation, i.e., the reconstructed velocity fields deviate much less from the DNS data compared to interpolated velocity fields.  Moreover, frequency spectra computed from the reconstructed velocity fields are found to closely resemble spectra obtained from the DNS data, even at frequencies higher than the Nyquist limit of the PIV-like data.  Note that the reconstruction framework developed here only makes use of 2D-2C velocity snapshots and simplified 
models grounded in the Navier-Stokes equations.  Unlike previous efforts \citep{tu2013integration,discetti2018estimation}, no additional time-resolved point measurements are used to improve temporal resolution. Instead, the observed improvement in temporal resolution stems from the use of RDT and TH to reconstruct the time evolution of the flow field from the spatial information present in the snapshots.

The remainder of this paper is structured as follows.  In Section~\ref{sec:methods}, we provide a brief review of the governing equations obtained under RDT and TH.  We also describe the methods used for \ml{generating the forward- and backward-time predictions}, the weighting functions used to fuse \ml{these} forward- and backward-time reconstructions of the flow field, and the metrics used for error quantification. In Section~\ref{sec:results}, we assess reconstruction accuracy for the various models developed in Section~\ref{sec:methods} (e.g., RDT vs. TH, different weighting functions).  We also evaluate the effects of measurement spatiotemporal resolution and noise on reconstruction accuracy, and compare turbulence statistics and spectra obtained from the reconstructed flow fields against DNS results.  Finally, we present concluding remarks in Section~\ref{sec:conclusions}.

\section{Methods}\label{sec:methods}
\subsection{Rapid Distortion Theory and Taylor's Hypothesis}
Under Rapid Distortion Theory, the 
Navier-Stokes equations are linearized about the mean profile \citep{batchelor1954effect, savill1987recent, hunt1990rapid} to yield the following momentum equation and continuity constraint:
\begin{equation}\label{eq:NSE}
\frac{\partial \ub}{\partial t} + \mathbf{U}\cdot \nabla \ub + \ub \cdot \nabla \mathbf{U} = -\nabla p + \frac{1}{Re_\tau}\nabla^2 \ub + (NL),
\end{equation}
and
\begin{equation}\label{eq:continuity}
\nabla \cdot \ub = 0.
\end{equation}
In the expressions above, $\mathbf{U}=(U(y),0,0)$ represents the mean profile, $\ub=(u,v,w)$ denotes the turbulent velocity fluctuations, $p$ represents pressure fluctuations, and $(NL)$ represents the (neglected) nonlinear terms.  A standard Cartesian coordinate system is used, in which $x$ is the streamwise direction, $y$ is the wall-normal direction, and $z$ is the spanwise direction; $t$ is time.  

Scaling arguments show that the nonlinear terms can be neglected in turbulent shear flows for time horizons that are shorter than the typical eddy turnover time \citep{savill1987recent, hunt1990rapid}. This makes RDT an appropriate choice for the present problem requiring temporal reconstruction between sequential PIV snapshots.  However, even with the substantial simplification afforded by linearization, reconstruction based on the full RDT equations is difficult in practice.  This is because most common PIV systems are only capable of generating planar 2D-2C field measurements.  Assuming these PIV measurements are carried out in the $(x,y)$ plane to yield velocity components $(u,v)$, additional simplifying assumptions are needed to account for the out-of-plane flow and pressure gradient terms.  Here, we simply neglect these terms to yield the following coupled advection-diffusion equations for streamwise and wall-normal velocity:
\begin{equation}
\underbrace{\frac{\partial u}{\partial t} +  U \frac{\partial u}{\partial x}}_{\text{advection term}} = \underbrace{\frac{1}{Re_\tau}\left(\frac{\partial^2 u}{\partial x^2} + \frac{\partial^2 u}{\partial y^2}\right)}_{\text{diffusion term}} -  \underbrace{v\frac{\partial U}{\partial y}}_{\text{coupling term}}\vc{,}
\label{eq:RDT_u}
\end{equation}
and
\begin{equation}
\underbrace{\frac{\partial v}{\partial t} +  U\frac{\partial v}{\partial x}}_{\text{advection term}} = \underbrace{\frac{1}{Re_\tau}\left(\frac{\partial^2 v}{\partial x^2} + \frac{\partial^2 v}{\partial y^2} \right)}_{\text{diffusion term}}. 
\label{eq:RDT_v}
\end{equation}
In other words, the continuity constraint is not enforced.  This ad hoc simplification is not rigorously justified.  However, solving two-dimensional versions of (\ref{eq:NSE}) and (\ref{eq:continuity}) also introduces additional modeling assumptions that are not rigorously justified and requires solution of the pressure Poisson equation. \ml{Moreover, the initial and final velocity fields do not satisfy the two-dimensional continuity equation.  So, imposing this constraint for the intermediate reconstructions could lead to additional numerical errors.}  Since the goal here is to generate simple 
models that can be used for flow reconstruction, we proceed with (\ref{eq:RDT_u})-(\ref{eq:RDT_v}).  \ml{These simplified, linearized versions of the Navier-Stokes equations are referred to as RDT for the remainder of this paper to acknowledge their conceptual origin (though we recognize that this terminology is not entirely accurate).}

Under Taylor's frozen turbulence hypothesis, the equations above are further simplified by assuming that the advection term is dominant. This yields
\begin{equation}\label{eq:TH}
\frac{\partial \ub}{\partial t} + U\frac{\partial \ub}{\partial x} = 0.
\end{equation}
As noted earlier, TH essentially assumes that the turbulent velocity field is `frozen' in form and advects downstream with the mean flow.  So, the local mean velocity can be used to convert between temporal variations and spatial variations. 

Below, we employ the linear models grounded in RDT (\ref{eq:RDT_u})-(\ref{eq:RDT_v}) and TH (\ref{eq:TH}) to reconstruct the time evolution of a turbulent channel flow in the time interval ($T$) between two planar 2D-2C field measurement snapshots (e.g., from PIV). \vc{Note that} the only input required for these reconstructions is the mean velocity profile \ml{appearing in (\ref{eq:RDT_u})-(\ref{eq:TH})}, which can be obtained from the snapshots themselves.

Keep in mind that reconstruction can proceed both forwards and backwards in time.  In other words, the equations above can be \ml{evolved} forwards in time using the first snapshot as the initial condition, as well as backwards in time using the second snapshot as the initial condition. \ml{These forward- and backward-time predictions can be generated via numerical integration of appropriately discretized versions of (\ref{eq:RDT_u})-(\ref{eq:TH}).  In addition, the advection equation in (\ref{eq:TH}) can also be evolved backwards or forwards in time using the method of characteristics.  When using discretized versions of the governing equations, the backward-time integration uses the transformation $\tau = T - t$, such that the new time variable has value $\tau = 0$ for the final snapshot at $t=T$, and value $\tau = T$ for the initial snapshot at $t=0$.  Note that this transformation switches the sign of the time derivative term in equations (\ref{eq:RDT_u})-(\ref{eq:TH}).  This is equivalent to solving the RDT equations with negative convection velocity, mean shear, and viscosity, and solving the TH equations with negative convection velocity}.  An appropriately weighted combination of these forward- and backward-time estimates has the potential to improve reconstruction accuracy.  Below, we develop physically motivated weighting schemes for this fusion of the forward- and backward-time estimates.  Of course, the time evolution of the flow between the two snapshots can also be estimated via linear interpolation, which does not require any underlying models or weighting schemes.  In Section~\ref{sec:results}, we show that the reconstruction framework developed in this paper yields velocity estimates that are significantly more accurate than estimates obtained via direct interpolation.

\subsection{Fusion of Forward and Backward Estimates}
The simplified linear equations obtained under RDT (\ref{eq:RDT_u})-(\ref{eq:RDT_v}) and TH (\ref{eq:TH}) are \ml{evolved} forwards in time from the first snapshot to yield a forward estimate for the velocity field, $\hat{\ub}_f$, and backwards in time from the second snapshot to yield a backward estimate, $\hat{\ub}_b$. The reconstructed flow field is given by a weighted average of these forward and backward estimates:
\begin{equation}
\hat{\ub} = k_{f} \hat{\ub}_f + k_{b} \hat{\ub}_b.
\end{equation}
A simple way to fuse the forward and backward estimates is to use weights that vary linearly in time
\begin{equation}\label{eq:weights}
k_{f} = k_f(t) = 1 - \frac{t}{T} \:\: ; \:\: k_{b} = k_b(t) = \frac{t}{T}.
\end{equation}
Here $t=0$ corresponds to the initial snapshot and $t=T$ corresponds to the final snapshot.  
This particular weighting scheme ensures that the forward-time estimate is weighted more heavily closer to the initial snapshot and the backward-time estimate is weighted more heavily towards the final snapshot.

\begin{figure}
    \centering
    \includegraphics[trim = {2cm 0cm 0cm 0cm},scale=0.575, clip=true]{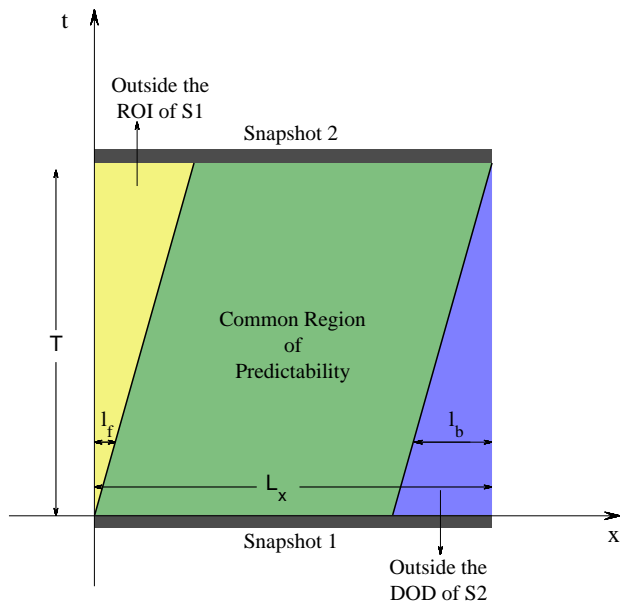}
    \caption{Schematic showing the combined Region of Influence (ROI) and Domain of Dependence (DOD) of two consecutive snapshots in the $x-t$ plane at a given wall-normal location.  Characteristics with slope determined by the local mean velocity ($dt/dx = 1/U(y)$) are shown by the solid black lines at the yellow-green and green-blue interfaces.  The green region shows where the ROI of Snapshot 1 (S1) and the DOD of Snapshot 2 (S2) coincide.  In this region both the forward and backward estimates are used for fusion. 
    }
    \label{fig:char_ROI}
\end{figure}

This weighting scheme can be improved further by considering the mathematical nature of the equations emerging from RDT and Taylor's hypothesis.  Since the hyperbolic advection term is expected to be dominant in wall-bounded turbulent flows, information is expected to propagate at a speed corresponding to the local mean velocity\ml{, $U(y)$}.  This is illustrated in the $x-t$ diagram shown in figure~\ref{fig:char_ROI}.  The region of influence (ROI) for the first snapshot and the domain of dependence (DOD) for the second snapshot are determined by characteristics in the $x-t$ plane that emanate from the upstream ($x=0$) and downstream ($x=L_x$) edge of the snapshots and have slope $dt/dx = 1/U(y)$. For advection-dominated flows, information propagation from the forward-time \ml{evolution} is confined to the ROI of the first snapshot, and information propagation from the backward-time \ml{evolution} is confined to the DOD of the second snapshot. In other words, the forward-time estimate is expected to be accurate only in the ROI of the first snapshot (i.e., green and blue regions in figure~\ref{fig:char_ROI}) while the backward-time estimate is expected to be accurate only in the DOD of the second snapshot (i.e., yellow and green regions in figure~\ref{fig:char_ROI}).  \ml{Further, since the slope of the characteristics that define the ROI and DOD is determined by the mean velocity, the size of the ROI and DOD also varies with $y$}.

To account for these effects, the weighting scheme in (\ref{eq:weights}) can be modified as follows.  The linear weighting scheme in (\ref{eq:weights}) can be retained in the common region of predictability for both snapshots (green region in figure~\ref{fig:char_ROI}).  The forward weight, $k_{f}$, is set to 0 in the region outside the ROI of the first snapshot (yellow region in figure~\ref{fig:char_ROI}) and the backward weight, $k_{b}$, is set to 1. Similarly, $k_{f}=1$ in the region outside the DOD of the second snapshot (blue region in figure~\ref{fig:char_ROI}) and $k_{b}=0$. The resulting equations for the weights are:
\begin{equation}
k_{f} = k_{f}(x,y,t) = \begin{cases}
0 & 0 \leq x<l_{f} \\
\left(1-\frac{t}{T}\right) & l_{f} \leq x \leq L_x - l_{b} \\
1 & L_x-l_{b}<x \leq L_x,
\end{cases}
\label{eq:weigtsKf_3D}
\end{equation}
and
\begin{equation}
k_{b} = k_{b}(x,y,t) = 
\begin{cases}
1 & 0 \leq x < l_{f} \\
\frac{t}{T} & l_{f} \leq x \leq L_x - l_{b} \\
0 & L_x-l_{b}<x \leq L_x.
\end{cases}
\label{eq:weigtsKb_3D}
\end{equation}
In the expressions above, $l_{f} = U(y)t$, $l_{b} = U(y)(T - t)$, and $L_x$ is the streamwise extent of the PIV window.  The upstream edge of the PIV window corresponds to $x=0$ and the downstream edge corresponds to $x=L_x$. Note that the weights shown in (\ref{eq:weigtsKf_3D})-(\ref{eq:weigtsKb_3D}) are dependent on $(x,y)$ as well as $t$, in contrast to temporal weighting scheme with $k_{f} = k_{f}(t)$ and $k_{b} = k_{b}(t)$ shown in (\ref{eq:weights}). \mkl{Moreover, notice that the weights are discontinuous at $x=l_{f}$ (yellow-green interface in figure~\ref{fig:char_ROI}) and at $x = L_{x} - l_{b}$ (green-blue interface in figure~\ref{fig:char_ROI}). Hence, this weighting scheme introduces spatial shocks in the reconstructed flow field at those locations.}

\subsection{Numerical Evaluation and Error Quantification}
To test reconstruction accuracy for the forward, backward, and fused flow field estimates, we use DNS data for turbulent channel flow at $Re_{\tau} = u_\tau h/\nu = h^+ = 1000$ available from the JHTDB \citep{graham2016web}.  Here, $h$ is the channel half-height, $u_\tau$ is the friction velocity, and $\nu$ is kinematic viscosity.  For consistency with basic PIV systems, we use 2D-2C velocity data in the $x-y$ plane that \vc{are} sampled uniformly in time and space (with one exception where the data are sampled with logarithmic spacing in the $y$-direction, as discussed below).  In other words, we use systematically sub-sampled DNS data as a surrogate for PIV measurements.  The results presented below make use of three complementary DNS datasets \ml{(see Table~\ref{tab:datasets})}.  For each dataset, the PIV window extends across the entire height of the channel ($2h$).  The streamwise extent is also set to $L_x = 2h$.

\begin{table}[]
\centering
\renewcommand{\arraystretch}{1.5}
\begin{tabular}{| c | c | c | c | c | c |}
\hline
Dataset & Grid Resolution & $\delta t^+$ & $N$ & Realizations & Section \\
\hline
1 & $\Delta x^{+} = \Delta y^{+} \approx 16$ & 0.0649 & 256 & 48 & III.A,D\\
\hline
2 & $\Delta x^{+} = \Delta y^{+} \approx 4$  & 0.0649 & 512 & 1 & III.B\\
\hline
3 & $\Delta x^{+} \approx 16$ & 0.649 & 1024 & 12 & III.C \\
 & Logarithmic in $y$ & & & & \\
\hline
\end{tabular}
\caption{\ml{DNS datasets acquired from the JHTDB \citep{graham2016web}. The time interval between individual snapshots in the dataset is $\delta t^+$ and the the total number of snapshots is $N$.  For datasets 1 and 3, multiple snapshot sequences are obtained for ensemble averaging purposes.  Dataset 1 includes 48 different realizations obtained at 6 different spatial locations and over 8 different time windows.  Dataset 3 includes 12 realizations obtained at different spatial locations. The last column lists the section(s) of the paper in which results corresponding to each dataset appear.}}
\label{tab:datasets}
\end{table}

\ml{Dataset 1 in Table~\ref{tab:datasets} is used to evaluate reconstruction accuracy for both RDT and TH for a benchmark test case}. This case has a uniform spatial grid comprising $129 \times 129$ points across the PIV window including the walls at the top and bottom.  
\ml{A total of $N=256$ snapshots are acquired at a sampling rate of $\delta t^+ = 0.0649$.}  Only the first and last snapshots in this dataset are used for reconstruction \ml{and so the prediction time horizon is $T^+ = N\delta t^+ \approx 16$}. The intervening snapshots from DNS are used to evaluate reconstruction accuracy. To evaluate statistical variations in reconstruction accuracy for this benchmark case, similar data are extracted at 6 different spatial locations and for 8 different time windows.  In other words, this dataset comprises 48 independent \ml{realizations} with identical spatiotemporal resolution and prediction time horizon.  Reconstruction accuracy for this benchmark case is discussed in Section~\ref{sec:accuracy}. 

\ml{Dataset 2 in Table~\ref{tab:datasets}} is acquired at higher spatial resolution with $513 \times 513$ uniformly sampled points across the $2h \times 2h$ PIV window.
This dataset includes \ml{$N=512$} snapshots obtained at intervals of $\delta t^+ = 0.0649$, for a total time horizon of $T^+ = N \delta t^+ \approx 33$.  In Section~\ref{sec:resolution}, this dataset is sub-sampled systematically to evaluate the effect of spatiotemporal resolution and prediction time horizon on reconstruction accuracy \vc{for both RDT and TH}. \ml{Dataset 3} is used to compute turbulence statistics and spectra in Section~\ref{sec:stats-spectra}.  For this, we use logarithmic spacing with $129$ grid points across the channel to better evaluate reconstruction accuracy in the near-wall region. In the streamwise direction, we use a uniform grid spacing of $\Delta x^+ \approx 16$, similar to the benchmark case.  For improved statistical convergence, a total of $N=1024$ snapshots are acquired \ml{at intervals of $\delta t^+ = 0.649$} over 12 different spatial locations of the DNS domain.  Reconstruction is carried out using every 25th snapshot in this dataset, so that the prediction time horizon $T^+ = 25\times \delta t^+ \approx 16$ is comparable to the benchmark case. \ml{Reconstruction is only carried out using the fused TH model for this dataset.}

As a point of comparison, for a physical system with PIV analysis being carried out for 16 pixel by 16 pixel segments with 50\% overlap (i.e., 8 pixels between data points), the benchmark case with $129 \times 129$ uniformly distributed points represents the use of a camera with approximately 1 Megapixel (1 MP) resolution.  The high-resolution dataset with $513 \times 513$  uniformly sampled points across the PIV window represents the use of a camera with 17 MP resolution.  Similarly, the time horizon for the benchmark case, $T^+ \approx 16$, corresponds to water flow with friction velocity $u_\tau = \sqrt{T^+ \nu / T} \approx 0.04$ ms$^{-1}$ for a PIV system capable of 100 Hz sampling rate ($T=0.01$ s) and $u_\tau \approx 0.15$ ms$^{-1}$ for a system capable of 1000 Hz sampling rate ($T = 0.001$ s).  For air flow, the corresponding friction velocity estimates are $u_\tau \approx 0.15$ ms$^{-1}$ for a 100 Hz system and $u_\tau \approx 0.5$ ms$^{-1}$ for a 1000 Hz system.  These estimates assume a kinematic viscosity of $\nu \approx 10^{-6}$ ms$^{-2}$ for water and $\nu \approx 1.5 \times 10^{-5}$ ms$^{-2}$ for air.  

For the reconstruction, \mkl{the simplified equations (i.e., (\ref{eq:RDT_u})-(\ref{eq:RDT_v}) for RDT and (\ref{eq:TH}) for TH)} are numerically integrated forwards in time from the first snapshot and backwards in time from the last snapshots over the prediction horizon.  A standard finite difference scheme is used for this purpose. An explicit Euler method is used for time integration, a first-order upwinding scheme is used for the advection terms, and a second-order central differencing scheme is used for the diffusion and coupling terms.  Numerical evaluation is carried out at the spatial resolution of the snapshots, and a time step of $\delta t^+ = 0.0649$.  The Courant–Friedrichs–Lewy (CFL) condition is satisfied for all parameter combinations. Due to the linear nature of the governing equations and the relatively short prediction horizons (512 time steps at most), the results presented below are not particularly sensitive to the choice of the numerical method. \ml{However, the finite difference discretization does introduce additional artificial viscosity.  The magnitude of this viscosity increases with increasing grid spacing \citep{roache1972artificial}.}

\ml{For TH, the effects of the artificial viscosity introduced by the numerical discretization of (\ref{eq:TH}) can be eliminated by using the method of characteristics to evolve the flow field in time. Specifically, the solution in the intervening period between the snapshots can be obtained by simply propagating the initial and final flow fields along characteristics determined by the mean velocity.  The forward-time evolution of the flow field can be computed from the initial snapshot $\ub\left(x,\hspace{2pt} y,\hspace{2pt} t=0\right)$ as:}
\begin{equation}\label{eq:char_f}
\ml{\hat{\ub}_f(x,y,t) = \ub\left(x-U(y)t,\hspace{2pt} y,\hspace{2pt} 0\right)}.
\end{equation}
\ml{Similarly, the backward-time evolution of the flow field can be computed from the final snapshot $\ub\left(x,\hspace{2pt} y,\hspace{2pt} t=T\right)$ as:}
\begin{equation}\label{eq:char_b}
\ml{\hat{\ub}_b(x,y,t) = \ub \left(x+U(y)(T-t),\hspace{2pt}y,\hspace{2pt}T\right)}.
\end{equation}
\ml{Here, $\hat{\ub}_f$ and $\hat{\ub}_b$ are the forward- and backward-time estimates and $U(y)$ is the mean velocity.  By construction, the forward-time estimate is confined to the ROI of the first snapshot ($l_f \leq x \leq L_{x}$ in figure~\ref{fig:char_ROI}), and the backward-time estimate is confined to the DOD of the second snapshot ($0 \leq x \leq L_{x}-l_{b}$ in figure~\ref{fig:char_ROI}).  The full reconstructed flow field can be obtained by fusing these forward- and backward-time estimates using the spatio-temporal weights shown in (\ref{eq:weigtsKf_3D})-(\ref{eq:weigtsKb_3D}).}

\begin{table}[]
\centering
\renewcommand{\arraystretch}{1.5}
\begin{tabular}{| c | c |}
\hline
Technique & Description \\
\hline
Interpolation & Linear interpolation in time \\ \hline
$RDT^{+}$ & Forward time integration of (\ref{eq:RDT_u})-(\ref{eq:RDT_v}) \\ \hline
$RDT^{-}$ & Backward time integration of (\ref{eq:RDT_u})-(\ref{eq:RDT_v}) \\ \hline
$RDT^{\pm}_{t}$ & Forward and backward time integration of (\ref{eq:RDT_u})-(\ref{eq:RDT_v}) \\
& Fused using temporal weights (\ref{eq:weights}) \\ \hline
$RDT^{\pm}_{tx}$ & Forward and backward time integration of (\ref{eq:RDT_u})-(\ref{eq:RDT_v}) \\
& Fused using spatio-temporal weights (\ref{eq:weigtsKf_3D})-(\ref{eq:weigtsKb_3D}) \\ \hline
$TH^{+}$ & Forward time integration of (\ref{eq:TH}) \\ \hline
$TH^{-}$ & Backward time integration of (\ref{eq:TH})\\ \hline
$TH^{\pm}_{t}$ & Forward and backward time integration of (\ref{eq:TH}) \\
& Fused using temporal weights (\ref{eq:weights}) \\ \hline
$TH^{\pm}_{tx}$ & Forward and backward time integration of (\ref{eq:TH}) \\
& Fused using spatio-temporal weights (\ref{eq:weigtsKf_3D})-(\ref{eq:weigtsKb_3D}) \\ \hline
$^*TH^{\pm}_{tx}$ & Characteristics-based evolution of (\ref{eq:TH}) \\
& Fused using spatio-temporal weights (\ref{eq:weigtsKf_3D})-(\ref{eq:weigtsKb_3D}) \\ \hline

\end{tabular}
\caption{\ml{Description of the different reconstruction techniques used in this study.  In general, superscripts ($+$, $-$, or $\pm$) represent time evolution while subscripts ($t$ or $tx$) represent the fusion scheme.  The $^*()$ notation denotes characteristics-based evolution of the flow field with TH.}}
\label{tab:techniques}
\end{table}

\ml{Table~\ref{tab:techniques} summarizes the different forward-time, backward-time, and fused reconstruction techniques used in this study.} Reconstruction accuracy is quantified using the following integrated error metrics.  The time-varying global error is defined as
\begin{equation}
\epsilon(t) = \frac{\left(\int_{x=0}^{2h}\int_{y=0}^{2h} \left(\left(u - \hat{u}\right)^2 + \left(v - \hat{v} \right)^2\right) dx dy\right)^\vc{{1/2}}}{\left(\int_{x=0}^{2h}\int_{y=0}^{2h}\left( u^2 + v^2\right) dx dy\right)^{\vc{1/2}}},
\label{eq:err}
\end{equation}
while the wall-normal variation in error over time is defined as
\begin{equation}
\epsilon(y,t) = \frac{\left(\int_{x=0}^{2h} \left(\left(u - \hat{u}\right)^2 + \left(v - \hat{v} \right)^2\right) dx\right)^\vc{{1/2}}}{\left(\int_{x=0}^{2h} \left(u^2 + v^2 \right) dx\right)^\vc{{1/2}}}.
\label{eq:err_y}
\end{equation}
In the expressions above, $\hat{u}$ and $\hat{v}$ are the reconstructed velocity fluctuations, $u$ and $v$ are the velocity fluctuations from DNS `truth', subsampled to match the PIV spatial resolution. The lower wall of the channel is located at $y=0$ and the upper wall is located at $y= 2h$. 
\section{Results and Discussion}\label{sec:results}
\ml{In this section, we evaluate reconstruction accuracy for all the techniques listed in Table~\ref{tab:techniques}}.
\ml{Reconstruction accuracy for the benchmark test case for the different models is evaluated in Section~\ref{sec:accuracy}.  \mkl{Specifically, we evaluate the time evolution of global error in Section~\ref{sec:accuracy_globErr}, the wall-normal variation in error over time in Section~\ref{sec:accuracy_wallErr}, and compare reconstructed flow fields with DNS snapshots in Section~\ref{sec:accuracy_comp}.} The effect of field measurement spatial resolution ($\Delta x^+ = \Delta y^+$) and time horizon ($T^+$) on reconstruction accuracy is considered in Section~\ref{sec:resolution}.  Reconstructed statistics and frequency spectra are shown in Section~\ref{sec:stats-spectra}.  This proof-of-concept study is based on field data from DNS. In Section~\ref{sec:noise}, we add Gaussian random noise to the first and last DNS snapshots used as inputs in the RDT/TH models to evaluate the effect of noisy real-world measurements on reconstruction accuracy. 
}

\subsection{Reconstruction Accuracy \vc{for Benchmark Test Case}}\label{sec:accuracy}

\subsubsection{\mkl{Variation in Global Error over Time}}\label{sec:accuracy_globErr}

\begin{figure}
    \centering
    \vspace{0.5cm}
    \hspace{3.5cm} \includegraphics[trim = {0cm 0cm 0cm 0cm},scale=0.750, clip=true]{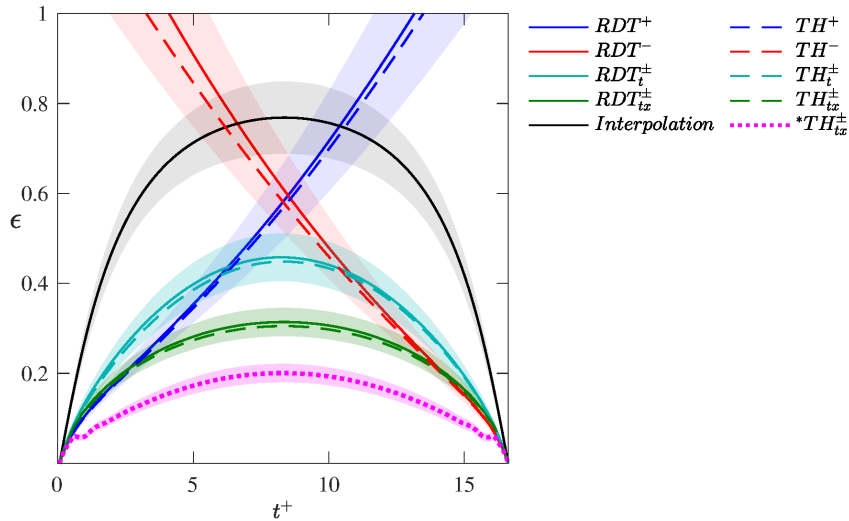}
    \caption{Time variation of integrated reconstruction error for the benchmark case with $T^+ \approx 16$ and $\Delta x^+ = \Delta y^+ \approx 16$. 
    The error is computed using (\ref{eq:err}) and averaged over 48 different \ml{realizations}. The solid, dashed, \vk{and dotted} lines represent the mean and the color bands are spaced at 1 standard deviation from the mean line.  \ml{Table~\ref{tab:techniques} provides a description of the different reconstruction techniques used in this figure}.
    }
    \label{fig:errorQuant}
\end{figure}

First, we compare reconstruction accuracy across different models for the benchmark case with spatial resolution $\Delta x^+ = \Delta y^+ \approx 16$ and prediction horizon $T^{+}\approx16$ \ml{using dataset 1 (see Table~\ref{tab:datasets})}.  The evolution in global error (\ref{eq:err}) for the simplest possible reconstruction technique --- linear interpolation between the snapshots --- is shown as the black line in figure~\ref{fig:errorQuant}.  Note that the error is averaged over 48 different \ml{spatiotemporal realizations} from the DNS database.  The mean error across these 48 different cases is plotted as a solid line and the shading represents one standard deviation about the mean.  As expected the error is zero at the beginning and the end of the prediction horizon, where snapshots of the flow field are available.  The error reaches a maximum of $\epsilon_{max} = \mathrm{max}(\epsilon)\approx0.8$ at the middle of the time horizon. 

Next, we evaluate reconstruction accuracy for the forward time integration of the RDT equations with the snapshot at $t^{+}=0$ used as the initial condition (denoted $RDT^+$).  With this technique, the global error is 0 initially and grows monotonically with time (solid blue line in figure~\ref{fig:errorQuant}).  The reconstruction error exceeds the error from linear interpolation at $t^{+} \approx 10$.  Similarly, the backward time RDT reconstruction ($RDT^-$) yields 0 error at the end of the time horizon (i.e., at $t^+ = T^+$ the initial condition for the backward time integration) and increases monotonically as time decreases.  The reconstruction error from $RDT^-$ exceeds that from linear interpolation for times before $t^+ \approx 6$.  In other words, for this case, the error dynamics are similar for both the forward and backward time RDT models.  

Note that linear interpolation outperforms the RDT-based models because it uses both the first and last snapshots over the prediction horizon for reconstruction.  In contrast, the forward RDT model uses only the first snapshot for reconstruction while the backward RDT model uses on the last snapshot. To improve reconstruction accuracy for the RDT-based models, we can fuse the forward and backward time estimate\vc{s} using the weighting functions shown in (\ref{eq:weights}) and (\ref{eq:weigtsKf_3D})-(\ref{eq:weigtsKb_3D}).  The fused RDT model that uses the temporal weighting function (\ref{eq:weights}) is denoted $RDT^{\pm}_t$.  The reconstruction error for this fused model (cyan line in figure~\ref{fig:errorQuant}) has a similar trend to the linear interpolation method, i.e., it has 0 error at the beginning and the end and reaches a maximum in the middle of the time domain.  However, the maximum error is significantly lower: $\epsilon_{max} \approx 0.5$ for $RDT^{\pm}_t$ compared to $\epsilon_{max} \approx 0.8$ for linear interpolation.  

Figure~\ref{fig:errorQuant} shows that the error from the fused estimate ($RDT^{\pm}_t$) exceeds that from forward and backward reconstructions alone at early and late times.  Ideally, any fusion would yield reconstructions that are as good as, or better than, the individual $RDT^+$ and $RDT^-$ estimates over the entire time horizon.  To a large extent, this can be achieved by taking into account the advection dominated nature of the flow under consideration.  As illustrated schematically in figure~\ref{fig:char_ROI}, the fused reconstruction can be further refined using the spatiotemporal weighting scheme given in equations~(\ref{eq:weigtsKf_3D})-(\ref{eq:weigtsKb_3D}).  This spatiotemporal reconstruction, termed $RDT^{\pm}_{tx}$, has the lowest maximum error of all the techniques considered thus far, with $\epsilon_{max} \approx 0.3$ (green line in figure~\ref{fig:errorQuant}).  Moreover, the $RDT^{\pm}_{tx}$ reconstruction yields errors comparable to $RDT^{+}$ and $RDT^{-}$ at early and late times, respectively. 

\ml{Next, we evaluate reconstruction accuracy for models grounded in Taylor's hypothesis (\ref{eq:TH}), and compare these reconstructions against those obtained using RDT. Similar to the RDT models, equation~(\ref{eq:TH}) can also be discretized and integrated forwards or backwards in time. These forward and backward time reconstructions under TH are denoted $TH^{+}$ and $TH^{-}$, respectively.  The forward and backward TH predictions can be combined using the temporal weights shown in (\ref{eq:weights}) to yield the fused estimate $TH^{\pm}_t$, or the spatiotemporal weights shown in (\ref{eq:weigtsKf_3D})-(\ref{eq:weigtsKb_3D}) to yield the fused estimate $TH^{\pm}_{tx}$.} \ml{The fused reconstruction obtained using the method of characteristics (\ref{eq:char_f})-(\ref{eq:char_b}) instead of numerical integration is denoted $^{*}TH^{\pm}_{tx}$.  Note that we only consider the fused $^{*}TH^{\pm}_{tx}$ reconstruction because the characteristics-based evolution of (\ref{eq:TH}) is not well defined outside the ROI of the first snapshot and the DOD of the second snapshot.} 

\ml{Figure~\ref{fig:errorQuant} shows the time evolution of error for the forward, backward, and fused reconstructions obtained using TH as \vk{dashed} lines.  When the discretized form of (\ref{eq:TH}) is used to generate the forward- and backward-time estimates, reconstruction accuracy for the TH models is nearly identical to that for the corresponding RDT models.  However, reconstruction performance improves further when the method of characteristics is used to evolve the flow field forwards and backwards in time \vk{as shown by the dotted magenta line}.  Maximum reconstruction error for $^*TH_{tx}^\pm$ is $\epsilon_{max} < 0.2$, compared to $\epsilon_{max} \approx 0.3$ for $TH_{tx}^\pm$ and $RDT_{tx}^\pm$.  This observation suggests that the artificial viscosity introduced by the numerical discretization of (\ref{eq:TH}) leads to a deterioration in reconstruction performance for the relatively coarse spatial resolution used in the benchmark test case ($\Delta x^+ = \Delta y^+ \approx 16$).  We explore this issue further in Sections~\ref{sec:resolution} and \ref{sec:stats-spectra}.  

Similar performance across the discretized RDT and TH models suggests that advection is the dominant physical mechanism retained in (\ref{eq:RDT_u})-(\ref{eq:RDT_v}).  The additional terms accounting for viscous effects and the interaction between wall-normal velocity fluctuations and mean shear appear to be less important.  However, keep in mind that the planar approximations to the RDT equations used here also neglect viscous effects due to out-of-plane gradients in the velocity, which are likely to be more important than the viscous effects arising due to streamwise gradients in velocity.  In other words, the neglected $\nu (\partial^2 \ub/\partial z^2)$ term is expected to be larger in magnitude than the  $\nu (\partial^2 \ub /\partial x^2)$ term.  Unfortunately, accounting for out-of-plane gradients in streamwise and wall-normal velocity is not possible with access to planar PIV measurements alone.  Moreover, as we show in Section~\ref{sec:stats-spectra}, reconstructions based on discretized versions of the governing equations underestimate the intensity of the wall-normal fluctuations in velocity. This may explain why inclusion of the additional $v (\partial U/ \partial y)$ term in (\ref{eq:RDT_u}) does not yield substantially different reconstructions relative to those based on TH.  
}

\begin{figure}
    \centering
    \includegraphics[trim = {0cm 0cm 0cm 0cm},width=0.9\textwidth, clip=true]{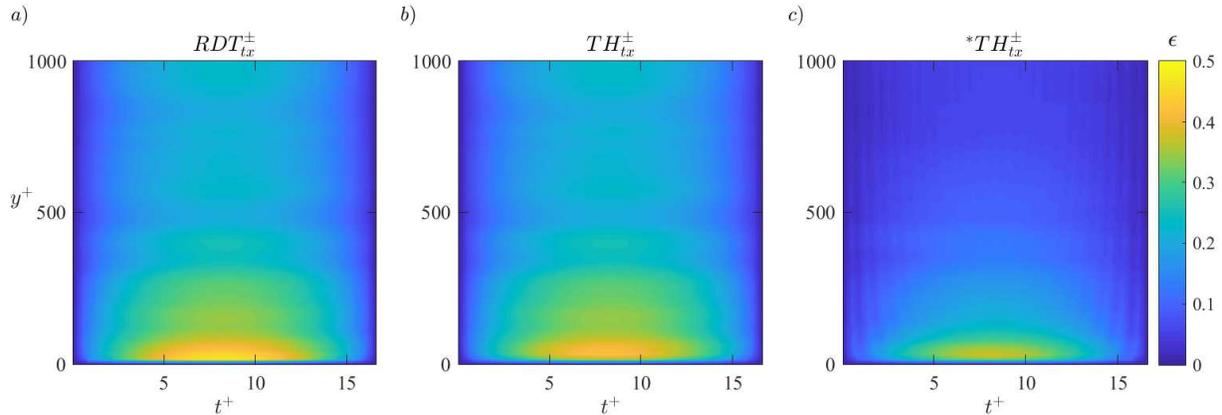}
    \caption{\ml{Wall-normal variation in reconstruction error as a function of time for the (a) $RDT_{tx}^{\pm}$ (b) $TH_{tx}^{\pm}$ and (c) $^*TH_{tx}^{\pm}$ models. The error is calculated using (\ref{eq:err_y}) and averaged over 48 different realizations.}
    }
    \label{fig:errorQuant_y}
\end{figure}

\begin{figure}
    \centering
    \includegraphics[trim = {0cm 0cm 0cm 0cm},width=0.8\textwidth, clip=true]{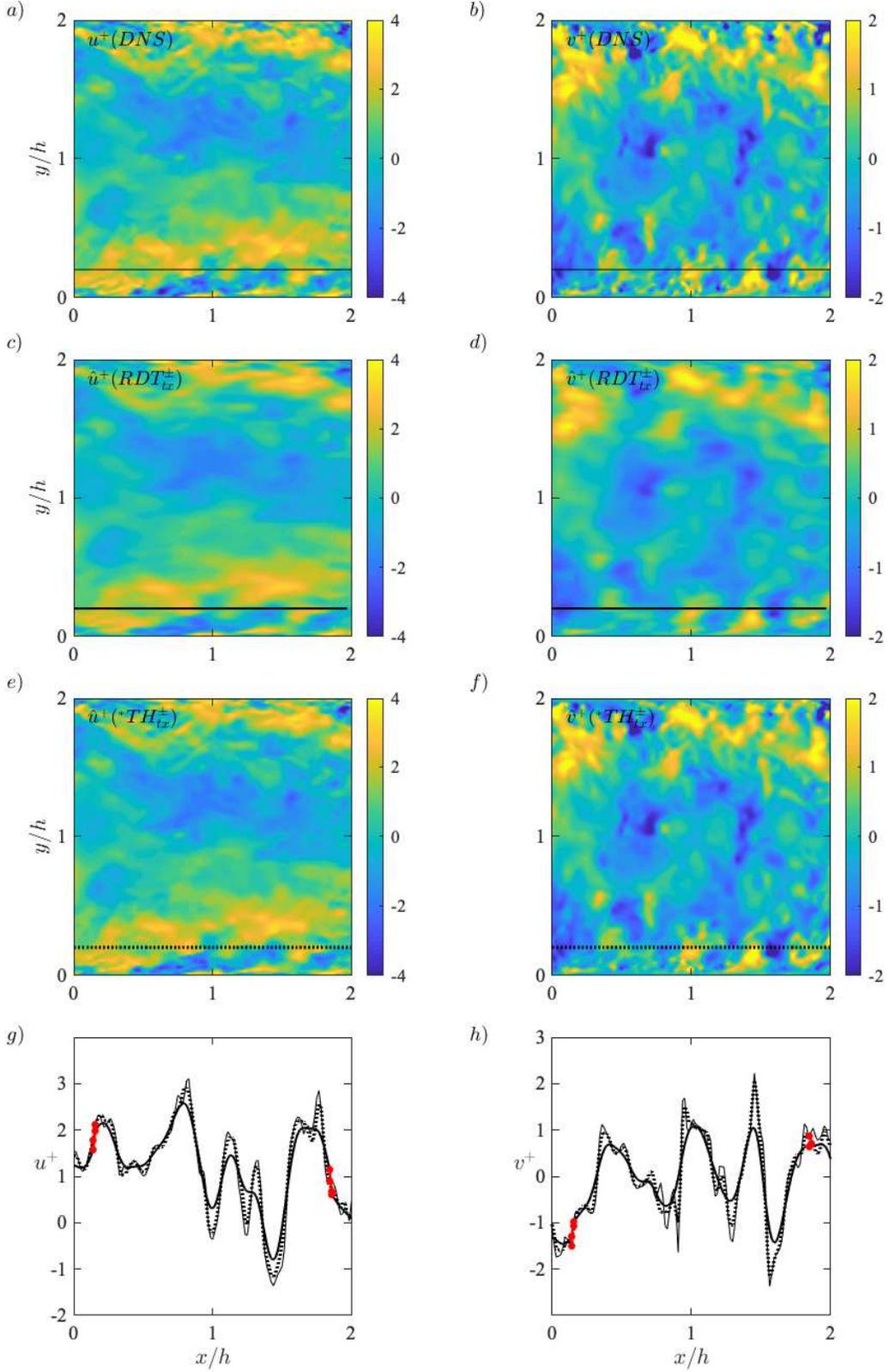}
    \caption{Snapshots of the velocity field from DNS (a,b)\ml{, the $RDT^{\pm}_{tx}$ reconstruction} (c,d)\ml{, and the $^{*}TH^{\pm}_{tx}$ reconstruction (e,f)} in the middle of the time horizon, when error is maximum.  
    Profiles of streamwise and wall-normal velocity at $y^{+}\approx 200$ are plotted in panels \vc{(g) and (h)}. The bold solid lines show the reconstructed velocity field \vc{for $RDT^{\pm}_{tx}$}, \vc{dotted lines are for $^{*}TH^{\pm}_{tx}$, and} the fine solid lines show the DNS velocity field. The locations of the shocks in the weighting functions for the reconstruction are highlighted in red.}
    \label{fig:x-tSnapshots}
\end{figure}

\subsubsection{\mkl{Wall-normal Variation in Error over Time}}\label{sec:accuracy_wallErr}

The time-evolution of error \ml{for the fused $RDT^{\pm}_{tx}$, $TH^{\pm}_{tx}$, and $^{*}TH^{\pm}_{tx}$ models} is shown as a function of wall-normal distance (\ref{eq:err_y}) in figure~\ref{fig:errorQuant_y}. Consistent with the plots in figure~\ref{fig:errorQuant}, the error is 0 at each $y$-location at the beginning and end of the time horizon and maximum in the middle \ml{for all models, and $^{*}TH_{tx}^{\pm}$ has the lowest reconstruction error at all wall-normal locations}. The error is in general higher in the inner region of the flow (below $y^+ \approx 200$ or $y/h \approx 0.2$) where turbulence production and turbulent kinetic energy are higher, and maximum reconstruction error increases closer to the wall. However, keep in mind that the first grid point is at $y^+ \approx 16$ due to the linear distribution of grid points.  This means that buffer region of the flow is not resolved completely. \ml{Figure~\ref{fig:errorQuant_y}(c) also shows the presence of distinct temporal oscillations in reconstruction error for $^*TH_{tx}^{\pm}$. The period for these oscillations decreases with increasing distance from the wall and closely matches the time scale $\Delta x^+/U^+$.  This suggests that the oscillations arise from specific grid locations leaving the ROI of the initial snapshot or entering the DOD of the later snapshot as time advances (see figure~\ref{fig:char_ROI}).}

The increase in reconstruction error with decreasing $y^+$ could be attributed to the reduction in turbulent timescales near the wall. Recall that \ml{linearization of the NSE is only} accurate for predictions over short periods of time. For $y^+ \approx 10$, the integral timescale for streamwise velocity fluctuations has been estimated to be $T_u^+ \approx 20$ while that for the wall-normal fluctuations has been estimated to be $T_v^+ < 10$ \citep{quadrio2003integral}.  Assuming these timescales are representative of typical eddy turnover times, the applicability of RDT \ml{and TH} is questionable in the near-wall region for the benchmark case with prediction horizon $T^+ \approx 16$.  Integral timescales for the velocity fluctuations are known to increase with distance from the wall \vc{and reach $T_u^+ \approx 110$ and $T_v^+ \approx 50$ at $y^{+} \approx 200$ \citep{choi2004lagrangian,luo2007lagrangian}}. So, the assumptions underlying \ml{RDT and TH} are better satisfied with increasing distance from the wall, which leads to an improvement in reconstruction accuracy.

\subsubsection{\mkl{Comparison of Reconstructed Flow Fields with DNS}}\label{sec:accuracy_comp}

To provide further insight into the reconstructed flow field\vc{s} obtained from the $RDT^{\pm}_{tx}$ \vc{and $^{*}TH^{\pm}_{tx}$} model\vc{s}, 
figure~\ref{fig:x-tSnapshots} compares spatial snapshots of the fluctuating velocity fields from DNS and the reconstruction\ml{s} in the middle of the reconstruction time horizon ($t/T = 0.5$), when the error is maximum. \vc{Panels (a), (c), and (e) show streamwise fluctuations from DNS, the $RDT^{\pm}_{tx}$ reconstruction, and the $^{*}TH^{\pm}_{tx}$ reconstruction, respectively. Similarly panels (b), (d), and (f) show wall-normal fluctuations from DNS, the $RDT^{\pm}_{tx}$ reconstruction, and the $^{*}TH^{\pm}_{tx}$ reconstruction, respectively.}
\vc{T}hese spatial snapshots show that the reconstructed velocity fields \ml{obtained from $RDT^{\pm}_{tx}$} qualitatively capture the large-scale structure.  However, they do not reproduce small-scale features of the turbulent flow field, particularly in the vicinity of the upper and lower walls. As shown in figures~\ref{fig:x-tSnapshots}\vc{(g)} and \ref{fig:x-tSnapshots}\vc{(h)}, the reconstructed flow fields \ml{for $RDT^{\pm}_{tx}$} appear to have gone through a \ml{spatial} low-pass filter when compared to the DNS. \ml{In contrast, reconstructed flow fields obtained from the $^{*}TH^{\pm}_{tx}$ model match the DNS results much more closely. This is also evident in figures~\ref{fig:x-tSnapshots}(g) and \ref{fig:x-tSnapshots}(h), which show that the $^*TH^{\pm}_{tx}$ profiles retain nearly all the small-scale features present in the DNS results, and there is minimal attenuation of fluctuation intensity.

Reconstructed flow fields obtained from the discretized $TH^{\pm}_{tx}$ model (not shown in figure~\ref{fig:x-tSnapshots}) closely resemble those obtained from the $RDT^{\pm}_{tx}$ model, i.e., they also appear low-pass filtered relative to DNS. This observation confirms that the artificial viscosity introduced by numerical discretization of the governing equations is responsible for the smoothing effect observed for the $RDT_{tx}^{\pm}$ and $TH_{tx}^{\pm}$ models.  Recall that these reconstructions are carried out using a much coarser spatial grid ($\Delta x^+ = \Delta y^+ \approx 16$; see Table~\ref{tab:datasets}) compared to DNS.  Since the artificial viscosity introduced by discretization is linearly proportional to grid resolution, we anticipate significant smoothing for the benchmark case.  The characteristics-based evolution of (\ref{eq:TH}) used for the $^*TH^{\pm}_{tx}$ reconstruction introduces no such artificial viscosity.}  


Finally, recall that the spatiotemporal weighting functions shown in (\ref{eq:weigtsKf_3D})-(\ref{eq:weigtsKb_3D}) have spatial shocks.  These spatial shocks are not visible in the reconstructed flow fields shown in figures~\ref{fig:x-tSnapshots}(c)\vc{-(f)}.  However, the streamwise profiles of velocity at $y^+ \approx 200$ shown in figures~\ref{fig:x-tSnapshots}(g) and \ref{fig:x-tSnapshots}(h) do show the presence of minor discontinuities in the reconstructed velocity field.  The locations of the shocks in the spatiotemporal weighting functions (\ref{eq:weigtsKf_3D})-(\ref{eq:weigtsKb_3D}) are highlighted in red for the reconstructed velocity profiles. The streamwise gradient in velocity is discontinuous at both ends of the red regions, but smooth elsewhere.  \ml{Together, figures~\ref{fig:errorQuant}-\ref{fig:x-tSnapshots} suggest that reconstruction accuracy is similar for the discretized RDT and TH models for the benchmark test case.  However, reconstruction accuracy improves further when the method of characteristics is used to generate TH-based reconstructions. Next, we assess the effect of measurement spatio-temporal resolution on reconstruction performance.}

\subsection{Effect of Measurement Spatiotemporal Resolution 
}\label{sec:resolution}

\begin{figure}
    \centering
    \includegraphics[trim = {0cm 0cm 0cm 0cm},width=0.9\textwidth, clip=true]{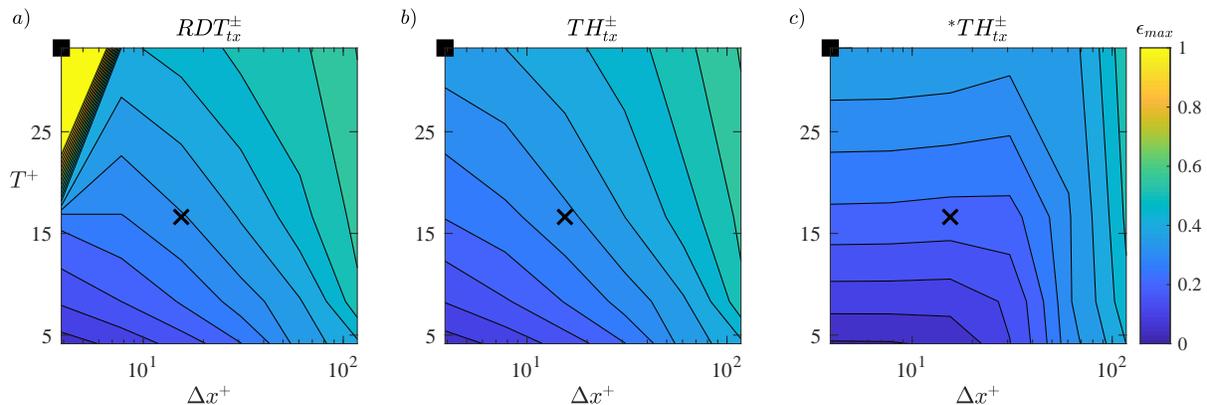}
    \caption{Maximum reconstruction error ($\epsilon_{max}$) as a function of the grid resolution ($\Delta x^{+}$) and time horizon $T^{+}$ \vc{for (a) $RDT_{tx}^{\pm}$,  (b) $TH_{tx}^{\pm}$, and (c) $^{*}TH^{\pm}_{tx}$.}  
    \ml{The contour lines are shown at intervals of 0.05, and}
    the benchmark case is shown as the black cross.  
    \ml{Figure \ref{fig:RDT_TH} shows the variation in integrated error as a function of time for the case marked with a black square ($\Delta x^{+} = 4; T^{+}=32$).} 
    }
    \label{fig:gridSens}
\end{figure}

In this section, we evaluate the accuracy of reconstruction as a function of spatial resolution and prediction time horizon \ml{using dataset 2 (see Table~\ref{tab:datasets})}. For the fused RDT \ml{and TH} model\ml{s} with spatiotemporal weights, the integrated error is 0 at the beginning and end of the prediction horizon, and reaches a maximum in the middle. This maximum error is plotted as a function of the grid resolution $\Delta x^{+} (= \Delta y^+)$ and time horizon $T^+$ in figures~\ref{fig:gridSens}\ml{(a), (b), and (c) for the $RDT^{\pm}_{tx}$, $TH^{\pm}_{tx}$, and $^*TH^{\pm}_{tx}$ reconstructions, respectively}. The grid resolution and time horizon for the benchmark case considered above are shown as a black cross.  

\ml{Consistent with the results shown in \mkl{Section~\ref{sec:accuracy_globErr}} for the benchmark case, reconstruction accuracy is broadly similar for the discretized forms $RDT^{\pm}_{tx}$ and $TH^{\pm}_{tx}$ models.  The characteristics-based reconstruction, $^{*}TH^{\pm}_{tx}$, shows similar trends, though reconstruction errors are generally lower and less sensitive to spatial resolution. For a given time horizon $T^{+}$, the maximum error increases gradually as a function of $\Delta x^+$ for the $RDT^{\pm}_{tx}$ and $TH^{\pm}_{tx}$ reconstructions. For the $^*TH^{\pm}_{tx}$ reconstruction, error is relatively insensitive to grid resolution below $\Delta x^+ \approx 20$ and only increases significantly beyond this threshold value for $\Delta x^+$. The gradual increase in reconstruction error as a function of $\Delta x^+$ for the $RDT^{\pm}_{tx}$ and $TH^{\pm}_{tx}$ models can be attributed to the artificial viscosity introduced by numerical discretization.  The magnitude of this artificial viscosity is expected to increase as a function of $\Delta x^+$ \citep{roache1972artificial}.  The $^*TH^{\pm}_{tx}$ reconstruction does not introduce any artificial viscosity.  However, for higher $\Delta x^+$, the initial and final snapshots include less information from smaller scale turbulent flow features. The forward and backward reconstructions are therefore unable to resolve these smaller scales, and reconstruction accuracy deteriorates.}

\begin{figure}
    \centering
    \includegraphics[trim = {0cm 0cm 0cm 0cm},width=0.9\textwidth, clip=true]{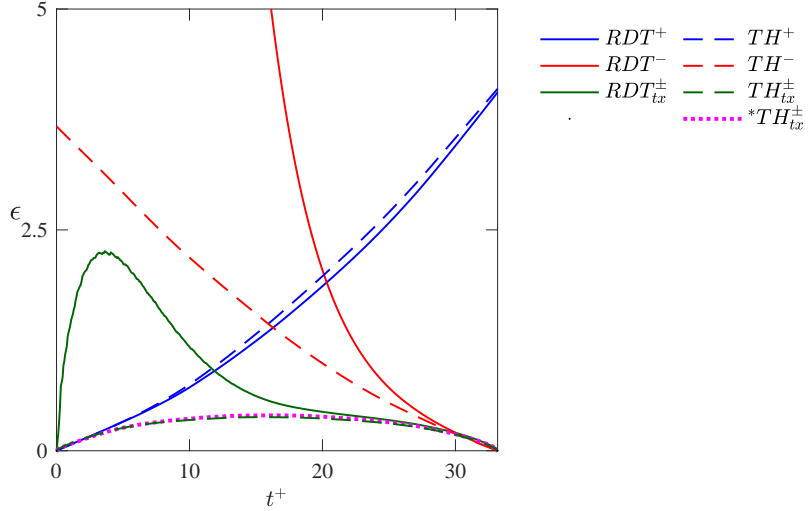}
    \caption{\ml{Variation in integrated error (\ref{eq:err}) as a function of time for the RDT, TH, and $^{*}$TH reconstructions at the grid resolution and time horizon marked with a black square ($\Delta x^{+} = 4; T^{+}=32$) in figure \ref{fig:gridSens}. 
    }
    }
    \label{fig:RDT_TH}
\end{figure}

For a given grid resolution $\Delta x^{+}$, the \ml{maximum reconstruction error increases as a function of time horizon for all the models.  As an example, for grid resolutions corresponding to the benchmark case ($\Delta x^+ = \Delta y^+ \approx 16$) the maximum error for $RDT_{tx}^{\pm}$ model increases from $\epsilon_{max} \approx 0.3$ for $T^+ \approx 16$ to $\epsilon_{max} \approx 0.5$ for $T^+ \approx 32$.} This observation is consistent with trends in figure~\ref{fig:RDT_TH}, which show that the forward and backward reconstruction errors increase monotonically with increasing and decreasing time, respectively. Hence, as prediction time horizon increases, any fusion of the forward and backward estimates is also expected to yield larger errors.  In general, as discussed earlier, any RDT- \ml{or TH}-based reconstructions \ml{are expected to} become less accurate as the prediction time horizon increases relative to typical turbulence timescales.

\ml{Notably, figure~\ref{fig:gridSens}(a) shows that} the error \ml{for the $RDT^{\pm}_{tx}$ reconstruction} grows dramatically with time horizon beyond $T^{+}=20$ at the lowest grid spacing considered here, $\Delta x^{+} \approx 4$.  \ml{In contrast, figures~\ref{fig:gridSens}(b) and \ref{fig:gridSens}(c) show no such increase in error for high grid resolutions and long prediction horizons for the $TH^{\pm}_{tx}$ and $^*TH^{\pm}_{tx}$ reconstructions. To provide further insight into this observation,} figure~\ref{fig:RDT_TH} shows the time evolution of error for \ml{all the RDT and TH} reconstructions for the case shown as a black square in figure~\ref{fig:gridSens}, \ml{which corresponds to} $\Delta x^+ \approx 4$ and $T^+ \approx 33$.  Compared to the forward RDT predictions (solid blue line), error for the backward RDT predictions (solid red line) grows much more steeply with time.  For instance, at $t^+ \approx 16$, the error associated with the forward RDT prediction is $\epsilon \approx 1$, while that for the backward RDT prediction is $\epsilon \approx 5$.  This blowup can be attributed to the viscous terms in the RDT equations (\ref{eq:RDT_u})-(\ref{eq:RDT_v}). \ml{As noted earlier,} when integrating backwards in time \ml{using the transformation $\tau = T - t$,} this viscous diffusion becomes negative.  This steepens spatial gradients in velocity and eventually leads to instability (especially if additional noise is introduced). In contrast, for the forward RDT integration, viscosity serves to smooth spatial gradients and damp external noise. The blowup in reconstruction error due to this negative diffusion phenomenon is most prominent for lower grid spacing and longer prediction horizons. This can be attributed to \ml{two reasons.  First, snapshots with finer grid resolutions are likely to include information from smaller-scale turbulent fluctuations with larger spatial gradients.  These large spatial gradients will be further amplified over time due to the negative effective viscosity in the backward RDT estimates.  Second,} the finite difference discretization scheme used here contributes additional artificial viscosity. \ml{The magnitude of this artificial viscosity is proportional to the grid spacing $\Delta x^{+}$ \citep{roache1972artificial}.} Hence, for higher grid spacing the effect of negative diffusion in the backward RDT estimates is partially offset by artificial viscosity. However, for grid spacing $\Delta x^+ \approx 4$, the effect of this artificial viscosity is outweighed by the negative diffusion over longer prediction horizons.

Neglecting the viscous diffusion terms in the RDT equations (\ref{eq:RDT_u}) and (\ref{eq:RDT_v}), along with the coupling term between horizontal and wall-normal velocity $v (\partial U/ \partial y)$ yields equation (\ref{eq:TH}), corresponding to Taylor's hypothesis. This is an advection equation for the velocity fluctuations with the mean velocity as the convection speed. \ml{This means that} the velocity fluctuations are advected downstream with \ml{speed $U(y)$ for the forward TH estimates and upstream with speed $U(y)$ for the backward TH estimates}.  The \vk{dashed} blue, red, and green lines in figure~\ref{fig:RDT_TH} show reconstruction accuracy as a function of time for the forward, backward, and fused TH models, respectively.  
When integrating forwards in time the performance of the TH model is comparable to that for the forward RDT model. However, when integrating backwards in time, the TH model far outperforms the RDT model.  While the backward RDT model led to $\epsilon \approx 5$ in the middle of the prediction horizon, the backward TH model yields $\epsilon \approx 1$.  This is comparable to the reconstruction error for the forward TH model, suggesting that the forward and backward time dynamics are similar under Taylor's hypothesis (as expected from the governing equations). \ml{Thus, eliminating the viscous diffusion term alleviates the sharp increase in reconstruction error observed for the backward RDT estimates. This also means that the fused $TH^{\pm}_{tx}$ model leads to significantly lower reconstruction error relative to the fused $RDT^{\pm}_{tx}$ model for fine spatial resolutions and longer prediction horizons (see solid and dashed green lines in figure~\ref{fig:RDT_TH}).} 

\ml{Note that reconstruction accuracy for the $^{*}TH_{tx}^{\pm}$ model is similar to that for the $TH_{tx}^{\pm}$ at the grid resolution considered in figure~\ref{fig:RDT_TH}. As mentioned above, the artificial viscosity introduced by numerical discretization increases as a function of grid spacing \citep{roache1972artificial}. Hence, at smaller grid resolutions both discretized and characteristics-based TH reconstructions yield similar results.} Together, these observations suggest that \ml{the $TH^{\pm}_{tx}$ and $^{*}TH_{tx}^{\pm}$ models}, which fuse the forward and backward TH estimates using the spatiotemporal weights shown in (\ref{eq:weigtsKf_3D})-(\ref{eq:weigtsKb_3D}), yield more robust and accurate reconstructions relative to the other techniques tested in this paper. 

\mkl{Keep in mind that the TH and $^*$TH models use spatial information to infer time evolution.  As a result, reconstruction accuracy is limited by the spatial resolution of the snapshots. The accuracy of the temporal reconstruction is expected to improve only if the frequency corresponding to the spatial Nyquist limit, $f_s^+ = U^+/ (2\Delta x^+)$, is higher than the temporal Nyquist frequency of the acquisition system, $f_t^+ = 1/(2T^+)$.  The expression for $f_s^+$ above translates the spatial Nyquist limit (i.e., only structures longer than $2\Delta x^+$ can be resolved) into a frequency using Taylor's hypothesis.  The requirement that $f_s^+$ be larger than $f_t^+$ translates into the following condition for spatial resolution: $\Delta x^+ < U^+ T^+$.}

For completeness, we note that discarding just the viscous terms from the RDT equations while retaining the coupling term in (\ref{eq:RDT_u}) does not yield any improvements in reconstruction accuracy relative to the TH models.  Similarly, \citet{yang2018implication} suggest the use of instantaneous streamwise velocity ($U+u$) instead of the the mean velocity in the near-wall region ($30\leq y^{+} \leq 100$) to improve estimates relative to Taylor's hypothesis.  For the cases considered here, this only leads to a marginal improvement in reconstruction accuracy ($< 0.5\%$).

\subsection{Reconstructed Statistics and Spectra}\label{sec:stats-spectra}

\begin{figure}
    \centering
    \includegraphics[trim = {0cm 0cm 0cm 0cm},width=0.9\textwidth, clip=true]{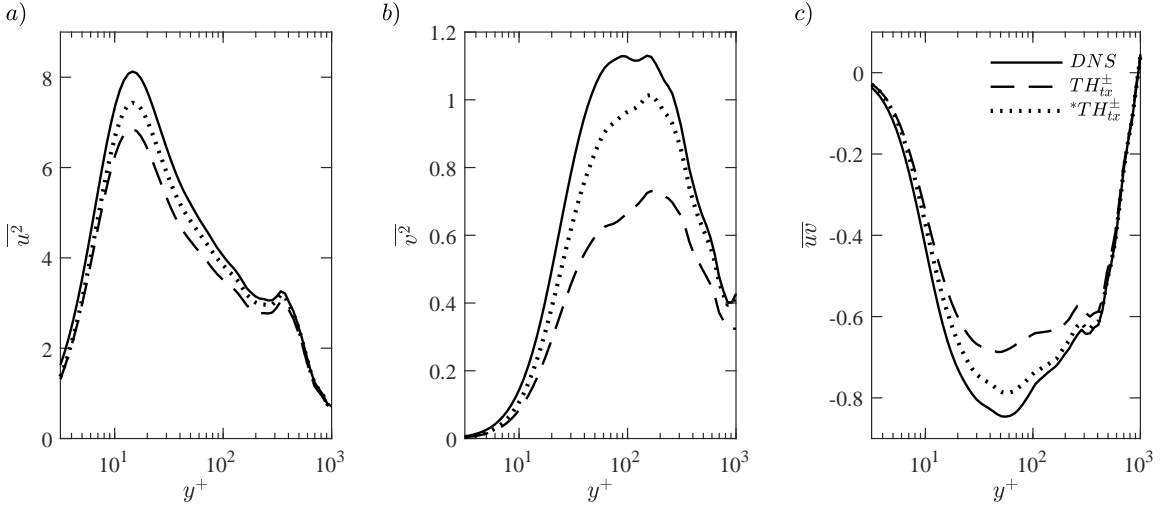}
    \caption{Comparison of wall-normal profiles of (a) $\overline{u^2}$, (b) $\overline{v^2}$, and (c) $\overline{uv}$ from DNS \ml{with the $TH^{\pm}_{tx}$ and $^{*}TH^{\pm}_{tx}$ reconstructions. The solid black lines show statistics computed directly from DNS data. The \vk{dashed} lines show statistics for $TH^{\pm}_{tx}$ and the \vk{dotted} lines show statistics for $^{*}TH^{\pm}_{tx}$}.  Reconstructions were carried out with the DNS data sub-sampled at intervals of $T^+ \approx 16$.}
    \label{fig:Stats}
\end{figure}

Results presented in the previous section show that the forward-time estimates of velocity obtained using TH are just as accurate as those obtained using RDT\ml{, when the governing equations are discretized}.  However, the backward-time estimates obtained using TH are more robust given the instability of the backward-time RDT predictions.  As a result, the fused $TH^\pm_{tx}$ model yields reconstructions that are just as accurate, and more robust, than the fused $RDT^{\pm}_{tx}$ model for the conditions tested in this paper. \ml{Reconstruction accuracy improves further when the TH reconstructions are obtained using a discretization-free method, at least for coarser grid resolutions.}  In this section, we compare single-point velocity statistics and frequency spectra obtained using the successful $TH^\pm_{tx}$ \ml{and $^{*}TH^\pm_{tx}$ models against DNS results.  We use dataset 3 for this evaluation (see Table~\ref{tab:datasets}).}

Figure~\ref{fig:Stats} compares profiles of the streamwise and wall-normal turbulence intensities ($\overline{u^2}$ and $\overline{v^2}$) as well as the Reynolds shear stress ($\overline{uv}$) obtained from DNS against those obtained from the $TH^{\pm}_{tx}$ \ml{and $^*TH^{\pm}_{tx}$} reconstruction\ml{s}. The prediction time horizon for these reconstructions, $T^+ \approx 16$, and the streamwise grid resolution, $\Delta x^+ \approx 16$, are identical to that for the benchmark case.  However, the grid points follow a logarithmic distribution in the wall-normal direction with a minimum spacing of ($\Delta y^+ \approx 3$) to allow for better comparison in the near-wall region. \mkl{As discussed in Section~\ref{sec:methods} only 41 sub-sampled DNS snapshots are used for reconstruction. The statistics are then computed for the reconstructed flow field.}  All reconstructed profiles show qualitative agreement with the DNS profiles. However, figure~\ref{fig:Stats}(a) shows that the near-wall peak in streamwise fluctuation intensity ($\overline{u^2}$) for the reconstructed flow field is lower than the DNS value by approximately 15\% \ml{for $TH^{\pm}_{tx}$ and 8\% for $^{*}TH^{\pm}_{tx}$}.  Figure~\ref{fig:Stats}(b) shows that the wall-normal fluctuation intensity ($\overline{v^2}$) is under-predicted to an even larger extent \ml{by the $TH^{\pm}_{tx}$ model}.  \ml{The peak value is reduced by approximately 35\% for $TH^{\pm}_{tx}$ and 10\% for $^{*}TH^{\pm}_{tx}$}.  Similarly, the Reynolds shear stress profile in figure~\ref{fig:Stats}(c) is under-predicted by about 20\% in the near-wall region \ml{for $TH^{\pm}_{tx}$ and 9\% for $^{*}TH^{\pm}_{tx}$}. 

\ml{Consistent with the results shown in figures~\ref{fig:errorQuant}-\ref{fig:gridSens}, the $^{*}TH_{tx}^{\pm}$ model outperforms the $TH_{tx}^{\pm}$ model and yields single-point statistics that are in close agreement with DNS data.  This improvement in performance for the $^{*}TH_{tx}^{\pm}$ model is illustrated well by the $\overline{v^2}$ profiles shown in figure~\ref{fig:Stats}(b). The $TH_{tx}^{\pm}$ reconstruction attenuates the $\overline{v^2}$ profile much more than the $^{*}TH_{tx}^{\pm}$ reconstruction.  This strong attenuation of the wall-normal velocity fluctuations is again caused by the artificial viscosity introduced by the numerical discretization of (\ref{eq:TH}). Since the wall-normal velocity fluctuations have more energy content at higher frequencies and smaller scales compared to the streamwise velocity fluctuations (see figure~\ref{fig:x-tSnapshots}), they are damped more quickly by the artificial viscosity. 

For completeness, we note that the $^*TH^{\pm}_{tx}$ reconstruction does attenuate the intensity and Reynolds' stress profiles by roughly $10\%$}. These errors would not be present in statistics computed from non-time resolved PIV snapshots, provided sufficient data are available to ensure convergence. Also keep in mind that the DNS profiles shown in figure~\ref{fig:Stats} are computed from a subset of the full database: \ml{$N = 1024$ snapshots obtained at intervals of $\delta t^+ \approx 0.649$ (i.e., total duration is $N\,\delta t^+ \approx 660$ viscous units)} for 12 different $2h \times 2h$ \ml{spatial} windows.  As a result, the profiles shown in figure~\ref{fig:Stats} are not expected to match the canonical converged profiles obtained from the full DNS.  The reconstruction is carried out with these data further sub-sampled in time: \ml{using every 25th snapshot such that the prediction horizon is similar to the benchmark case, $T^+ = 25\delta t^+ \approx 16$}.

\begin{figure}
    \centering
    \includegraphics[trim = {0cm 0cm 0cm 0cm},width=0.9\textwidth, clip=true]{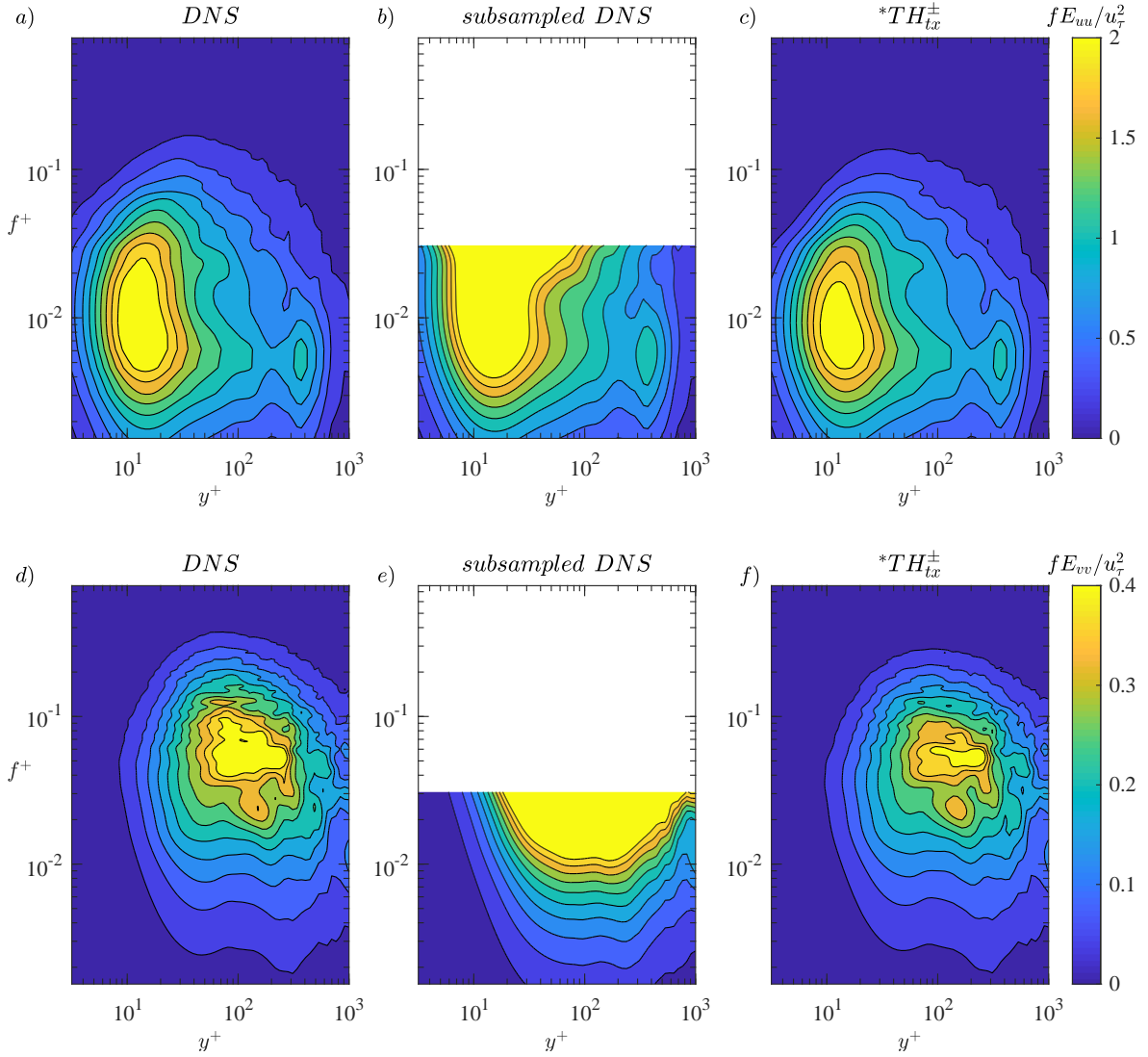}
    \caption{Premultiplied power spectral density for streamwise velocity $f E_{uu}/u_\tau^2$, \vc{(a) - (c)},  \vc{and  wall-normal velocities $f E_{vv}/u_\tau^2$, (d) - (f).} 
    Panels \vc{(a) and (d) are} obtained from DNS data. Panels (b) \vc{and (e) show the} \vc{p}ower spectra from the PIV-like data, i.e., DNS data sub-sampled \ml{at a frame rate corresponding to $T^+ \approx 16$}. 
    Panels (c) \vc{and (f) show the} \vc{p}ower spectra obtained from the \ml{characteristics-based $^{*}TH^{\pm}_{tx}$ reconstruction. Contour lines are shown at intervals of 0.2 for panels (a)-(c), and 0.04 for panels (d)-(f)}.  
    }
    \label{fig:SpectraEuuEvv}
\end{figure}

The premultiplied power spectral density for the streamwise and wall-normal velocity fluctuations \mkl{are} shown in figure~\ref{fig:SpectraEuuEvv}, plotted as a function of frequency and wall-normal location.  Panel\vc{s} (a) \vc{and (d)} show results obtained from DNS data. Panel\vc{s} (b) \vc{and (e)} show results obtained from the PIV-like data i.e., DNS data sub-sampled at intervals of $T^{+} \approx 16$.  Panel\vc{s} (c) \vc{and (f)} show results computed from the reconstructed velocity fields \ml{obtained using the best performing $^{*}TH_{tx}^{\pm}$ model}.  The classical inner peak at $y^+ \approx 15$ and $f^{+} \approx 10^{-2}$ from the near-wall cycle \citep{smits2011high}, which corresponds to streak-like structures with a streamwise wavelength of $\lambda_x^+ = U^{+}(y^{+} \approx 15)/f^{+} \approx 10^3$, is clearly evident in the DNS data for streamwise velocity \ml{shown in figure~\ref{fig:SpectraEuuEvv}(a)}.  However, figure~\ref{fig:SpectraEuuEvv}(b) shows that this peak is not resolved in the sub-sampled PIV-like data. This is a direct consequence of Nyquist's sampling criterion. For the PIV-like data, frequencies higher than $f^+ = 0.5/T^+$ are not resolved.  This translates into $f^{+} \approx 3\times10^{-2}$ for $T^+ \approx 16$, which is insufficient to fully resolve the near-wall peak.  As a result, the distinct peak in the DNS data around $f^{+}\approx 10^{-2}$ is replaced by a region of high power spectral density at lower frequencies \ml{because of aliasing}.  In contrast, figure~\ref{fig:SpectraEuuEvv}(c) shows that the premultiplied spectra computed \ml{from the $^{*}TH^{\pm}_{tx}$ reconstruction} closely match those in figure~\ref{fig:SpectraEuuEvv}(a), both in terms of near-wall peak location and shape.  The magnitude of the peak is \ml{slightly} reduced for the reconstruction relative to that \vc{of} the DNS data, which is consistent with the \ml{reduction in the magnitude of the reconstructed $\overline{u^2}$ profile observed in figure~\ref{fig:Stats}(a)}. The wall-normal velocity spectra in figures~\vc{\ref{fig:SpectraEuuEvv}(d) - \ref{fig:SpectraEuuEvv}(f)} show a similar trend.  \ml{For the wall-normal velocity fluctuations, the DNS spectra in figure~\ref{fig:SpectraEuuEvv}(d) show a peak centered around wall-normal location $y^+ \approx 100$ and frequency $f^+ \approx 7\times 10^{-2}$, which is above the Nyquist limit of the PIV-like sub-sampled data.  The reconstructed spectra shown in figure~\ref{fig:SpectraEuuEvv}(f) capture the location of this peak reasonably well.} The magnitude of the power spectral density is \ml{again under-predicted slightly,}
\ml{which is consistent with the reduction in magnitude of the $\overline{v^2}$ profile observed in figure~\ref{fig:Stats}(b) for the reconstruction}.

\begin{figure}
    \centering
    \includegraphics[trim = {0cm 0cm 0cm 0cm},width=0.9\textwidth, clip=true]{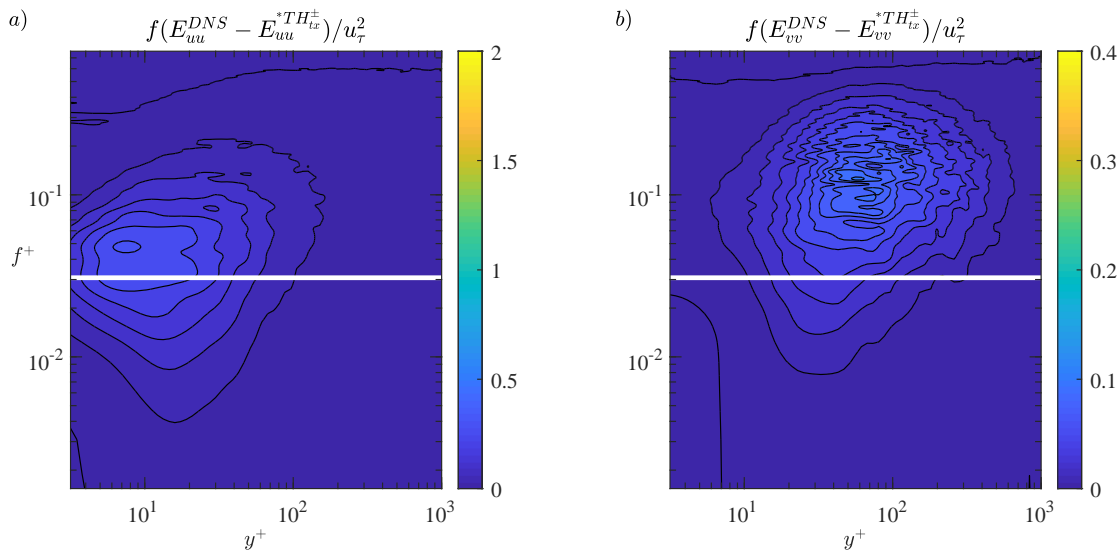}
    \caption{\vc{Error in \ml{reconstructed} power spectral densities for (a) streamwise velocity, and (b) wall-normal velocity. The error is computed \ml{by taking the difference between} power spectra obtained from DNS and the \ml{$^{*}TH^{\pm}_{tx}$} reconstruction. The horizontal white line corresponds to the Nyquist limit for the PIV-like data sampled at $T^+ \approx 16$. Contour lines are shown at intervals of 0.05 for (a), and 0.01 for (b).}}
    \label{fig:errSpectra}
\end{figure}

\ml{Figure~\ref{fig:errSpectra} shows the error in the reconstructed power spectral densities relative to DNS for both the streamwise velocity fluctuations, $f (E_{uu}^{DNS} - E_{uu}^{^{*}TH_{tx}^{\pm}})/u_\tau^2$, and the wall-normal velocity fluctuations, $f (E_{vv}^{DNS} - E_{vv}^{^{*}TH_{tx}^{\pm}})/u_\tau^2$. As expected, for both components of velocity, reconstruction errors are generally largest in the vicinity of the peaks in the spectra.  It is also clear that errors are larger for frequencies above the Nyquist limit for the sub-sampled PIV-like data used for the reconstruction.  Consistent with the results shown in figure~\ref{fig:SpectraEuuEvv}, the reconstructed velocity spectra deviate very little from the DNS results below the Nyquist limit.  In contrast, the spectra obtained directly from the PIV-like snapshots (figures~\ref{fig:SpectraEuuEvv}(b),(e)) deviate substantially from the DNS results.  In other words, the reconstruction procedure used here improves spectral predictions even below the temporal Nyquist limit of the input data. The ability to effectively reproduce these lower-frequency fluctuations arising from larger-scale turbulent flow structures is likely to become even more important in the reconstruction of wall-bounded turbulent flows at higher Reynolds numbers. The fact that the $^{*}TH^{\pm}_{tx}$ model is also able to reproduce important features in the frequency spectra above the temporal Nyquist limit of the input data is an added bonus.}

As noted earlier, Taylor's hypothesis has long been used to convert premultiplied frequency spectra, such as those shown in figure~\ref{fig:SpectraEuuEvv} but obtained from time-resolved point measurements, into estimates of wavelength spectra.  The time-resolved nature of the point measurements ensures that features with small streamwise wavelengths, which would appear as higher frequencies per Taylor's hypothesis ($f^+ = U^+/\lambda_x^+$), are resolved.  Here, we use TH to reconstruct the time evolution of the flow field from non-time resolved field measurements.  In this case, the small-scale spatial information present in the intermittent velocity snapshots enables us to resolve dynamics at frequencies higher than the temporal Nyquist limit.  In other words, since the PIV-like measurements are able to resolve flow structures with streamwise wavelengths as small as $2\Delta x^+$, this spatial information can be used to estimate spectra for frequencies up to $f^+ = U^+/(2\Delta x^+)$ under TH.


\subsection{Effect of Noise on Reconstruction Accuracy}\label{sec:noise}

\begin{figure}
    \centering
    \includegraphics[trim = {0cm 0cm 0cm 0cm},width=0.9\textwidth, clip=true]{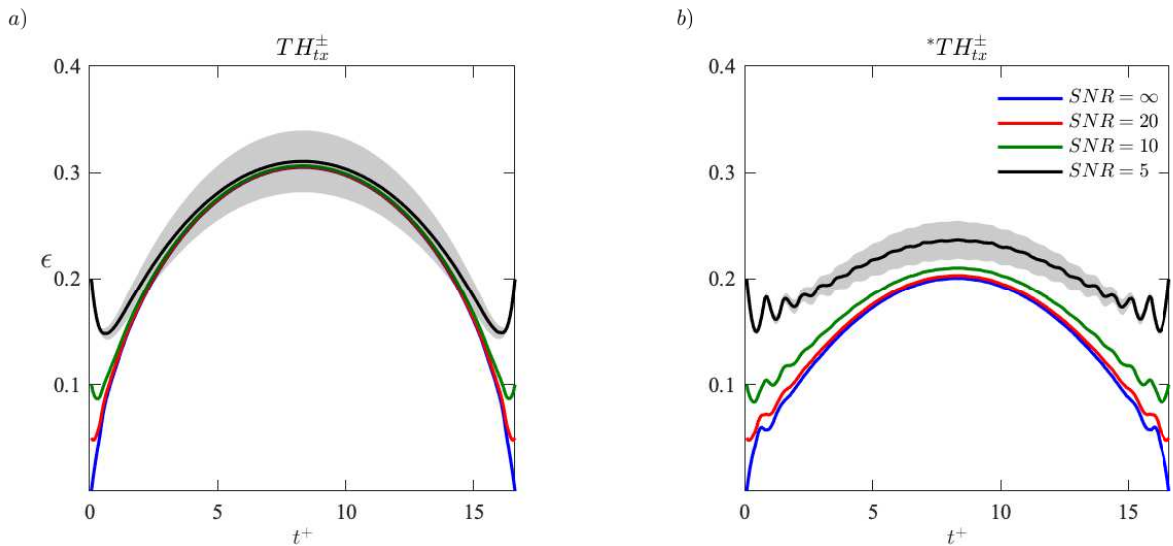}
    \caption{\ml{Variation of integrated error as a function of time with 4 different initial noise levels for (a) $TH^{\pm}_{tx}$ and (b) $^*TH^{\pm}_{tx}$.} The solid lines show results averaged over 48 different realizations. A shaded gray band representing one standard deviation above and below the average value from the 48 realizations is shown for the noisiest case, i.e., $SNR=5$. 
    }
    \label{fig:errQuantNoisy}
\end{figure}

Finally, we briefly evaluate the effect of random measurement noise on reconstruction accuracy \ml{for the} $TH^{\pm}_{tx}$\ml{, and $^{*}TH^{\pm}_{tx}$ models}.  Specifically, we add zero-mean, independent and identically distributed Gaussian random noise of varying intensity to the initial and final snapshots prior to reconstruction.  Four different signal to noise ratios are tested, $SNR = (\infty, 20, 10,5)$. $SNR=\infty$ corresponds to the noise free DNS data tested thus far. Figure~\ref{fig:errQuantNoisy} shows reconstruction accuracy as a function of time for the benchmark case \ml{for both models}.  Similar to figure~\ref{fig:errorQuant}, these results are averaged over \ml{the} 8 different temporal and 6 different spatial windows \ml{included in dataset 1} (i.e., 48 different realizations) for each noise level.  At the beginning and end of the prediction horizons, the integrated error corresponds to the level of noise added.  For example, the case with $SNR = 5$ leads to $\epsilon = 0.2$ for the initial and final snapshots.  

\ml{For the $TH^{\pm}_{tx}$ reconstructions shown in figure~\ref{fig:errQuantNoisy}(a), this added noise gets attenuated quickly over time}.  In the middle of the prediction horizon, where the error is maximum, reconstruction accuracy for all four $SNR$ values is similar.  In other words, reconstruction accuracy for the noisiest snapshots ($SNR =5$, black line) is very close to that for the noise-free data ($SNR = \infty$, blue line). Averaged over the 48 different realizations, there is only a difference of $1.8\%$ in $\epsilon_{max}$ between the noise-free case and the noisiest case with $20\%$ initial error.  The initial attenuation of noise can be attributed to the artificial viscosity introduced by the finite difference discretization used here.  This artificial viscosity attenuates any spatial gradients in velocity introduced by the random noise. \ml{The $^{*}TH_{tx}^{\pm}$ reconstructions in figure~\ref{fig:errQuantNoisy}(b) do not show as much damping of the initial noise. The maximum reconstruction error, $\epsilon_{max}$, is approximately 4\% larger for the noisiest case with $SNR=5$ compared to the noise-free case.  Despite the reduced damping, the $^{*}TH_{tx}^{\pm}$ model yields more accurate reconstructions than the $TH_{tx}^{\pm}$ model for all noise levels.

Note that the $^{*}TH_{tx}^{\pm}$ reconstructions show distinct temporal oscillations in error. These oscillations are most evident for the noisiest case with $SNR=5$.  When the error evolution is evaluated using (\ref{eq:err_y}) for a particular $y$ location, the oscillation period closely matches the time scale $\Delta x^{+}/U^{+}(y)$.  This indicates that the oscillations arise from individual grid points leaving the ROI of the first snapshot or entering the DOD of the subsequent snapshot as time advances.  Similar oscillations in reconstruction error are also observed for the $TH^{\pm}_{tx}$ model. However, these oscillations are masked by the larger magnitude of the error.  Smoothing due to artificial viscosity may also play a role.}

Real-world measurements are likely to suffer from both random and systematic error.  Random errors can be modelled reasonably using Gaussian white noise, as is done here.  However, accounting for systematic errors due to hardware limitations or analysis procedures requires different models (e.g., multiplicative or additive noise of varying intensity).  A detailed evaluation of such errors is outside the scope of the present effort. 

\section{Conclusions}\label{sec:conclusions}
The results presented in this paper show that both RDT and Taylor's hypothesis can lead to useful reconstructions of wall-bounded turbulent flows from non-time resolved PIV snapshots. Compared to previous reconstruction efforts, the methods proposed here are distinguished by the following features.  First, the methods are based on the governing Navier-Stokes equations, with  associated simplifying assumptions.  Second, the models use spatial information from the snapshots directly to infer time evolution; no additional time-resolved measurements are needed.  Third, we evolve the flow fields both forwards and backwards in time, and fuse these estimates to improve reconstruction accuracy.  This fusion is carried out using spatiotemporal weighting functions that also take advantage of the advection dominated nature of wall-bounded turbulent flows (figure~\ref{fig:char_ROI}). \ml{The only input required for these reconstructions is the mean velocity profile appearing in equations (\ref{eq:RDT_u})-(\ref{eq:TH}), which can be obtained from the PIV data.}

Overall, the use of Taylor's hypothesis, with the local mean velocity ($U$) as the convection speed, leads to more robust and accurate reconstructions compared to the use of models grounded in RDT.  This result is somewhat counter-intuitive since the RDT equations (\ref{eq:RDT_u})-(\ref{eq:RDT_v}) include additional flow physics. The forward time estimates from the RDT and TH models are of comparable accuracy. However, the backward estimates from RDT are prone to instability due to the negative diffusion introduced by backward time integration. This leads to substantial reconstruction errors over long time horizons, especially from datasets with high spatial resolution (figure~\ref{fig:gridSens}). \ml{The accuracy of the TH reconstructions improves further when the method of characteristics is used to evolve the flow field instead of time integrating the discretized form of (\ref{eq:TH}). This is because numerical discretization introduces artificial viscosity, which serves to damp out smaller-scale features present in the initial and final snapshots.} Consistent with physical intuition, reconstruction accuracy using the fused TH model improves as the spatial resolution of the snapshots improves and as the prediction time horizon between snapshots gets smaller. Using the instantaneous streamwise velocity ($U+u$) in Taylor's hypothesis compared to the mean streamwise velocity does not yield a substantial improvement in performance.  Since the convection speed of near-wall turbulent flow structures is known to converge to $c^+ \approx 10$ below $y^+ \approx 15$ \mkl{\citep{delAlamo2009estimation}}, perhaps reconstruction accuracy can be improved further by altering the convection velocity in the viscous sub-layer and buffer region of the flow. 

The fused TH model \ml{that utilizes the method of characteristics to evolve the flow field} also proves to be useful in reconstructing premultiplied spectra for frequencies that are above the Nyquist limit of the acquisition rate of the PIV-like data. Spectra computed using the flow fields reconstructed from DNS data sub-sampled significantly in time closely resemble spectra computed directly from the DNS data. 
The reconstruction is also robust to random external noise, as shown in figure~\ref{fig:errQuantNoisy}. The effect of systematic errors in the field measurements remains to be studied. 

High Reynolds number wall-bounded turbulent flows are known to be advection dominated. Hence, the success of Taylor's hypothesis and the spatiotemporal weighting scheme depicted in figure~\ref{fig:char_ROI} in flow reconstruction is perhaps not surprising. Taylor's hypothesis has been used extensively in previous studies to extract spatial information from temporal measurements \citep[e.g.,][]{wyngaard1977taylor,dennis2008limitations,moin2009revisiting}. However, in this study, we use spatial information from field measurements to infer the time evolution of the flow between two consecutive PIV-like snapshots. \mkl{As discussed in Section~\ref{sec:resolution},} since spatial information is used to infer time evolution, the resolution of the spatial data limits reconstruction accuracy.  \mkl{The accuracy of the temporal reconstruction is expected to improve only if the frequency corresponding to the spatial Nyquist limit is higher than than the frequency corresponding to the temporal Nyquist limit of the acquisition system. This requirement translates into the following condition for spatial resolution $\Delta x^+ < U^+ T^+$.} 
The other limit on reconstruction is imposed by the \ml{hyperbolic nature of the governing equations for both RDT and TH.  Equations (\ref{eq:RDT_u})-(\ref{eq:RDT_v}) or (\ref{eq:TH})}, can only be used to accurately reconstruct the flow field if the advection time scale $L_x^{+}/U^+$, where $L_x^+$ is the streamwise extent of the snapshot, is less than the prediction time horizon, $T^+$. If this condition is not met, there will be regions in the $x-t$ plane that are not covered either by the region of influence for the first snapshot or the domain of dependence for the last snapshot.
\begin{acknowledgments}
CVK gratefully acknowledges the graduate school at University of Southern California for financial support through the Provost fellowship.
\end{acknowledgments}
\bibliographystyle{unsrtnat}

\end{document}